\documentclass[longauth]{aa} 
\usepackage{longtable,lscape}
\usepackage{natbib}
\usepackage{graphicx}
\usepackage{amsmath}
\usepackage[colorlinks=true,citecolor=blue]{hyperref}
\usepackage{xcolor}
\usepackage{amssymb}
\usepackage{txfonts}

\def \m2s2{\,m$^{2}$\,s$^{-2}$} 
\def \kms{\,km\,s$^{-1}$}       
\def \vsini{$v\sin{i}$}         
\def \msini{$m\sin{i}$}         
\def \Msol{M_\odot}             
\def \Msun{M_\odot}             
\def \Mjup{M_{\text{Jup}}}
\def \Rjup{R_{\text{Jup}}}
\def \banyan{{\tt BANYAN} $\Sigma$}
\def \teff{$T_{\text{eff}}$}
\def \snr{\text{S/N}}
\def \ruwe{\text{RUWE}}
\def \pma{\text{PMa}}
\def \prot{$P_{\text{rot}}$}
\def \snrruwe{(\snr)_\ruwe}
\def \snrpma{(\snr)_\pma}

\defcitealias{desidera21}{D21}
\defcitealias{chauvin26}{Chauvin et al. in prep.}

\usepackage{upgreek}

\begin{document}
\title{The SPHERE infrared survey for exoplanets (SHINE)}
\subtitle{V. Full sample characterization}

\author{
V.~Squicciarini \inst{\ref{pd},\ref{exeter}} \and
S.~Desidera \inst{\ref{pd}} \and
G.~Chauvin \inst{\ref{mpia},\ref{oca}} \and
F.~Kiefer \inst{\ref{lira}} \and
V.~D'Orazi\inst{\ref{roma},\ref{pd},\ref{austin}} \and
C.~Fontanive\inst{\ref{edinb}} \and
A.~Vigan\inst{\ref{lam}} \and 
D.~Nardiello \inst{\ref{pd}} \and
S.~Messina\inst{\ref{ct}} \and
D.~Albert \inst{\ref{osug}} \and
S.~Bergeon \inst{\ref{ipag}} \and
J.-L.~Beuzit\inst{\ref{ipag},\ref{lam}} \and
B.~Biller\inst{\ref{edinb}} \and 
A.~Boccaletti\inst{\ref{lira}} \and
M.~Bonavita\inst{\ref{pd}} \and
M.~Bonnefoy\inst{\ref{ipag}} \and
W.~Brandner\inst{\ref{mpia}} \and
F.~Cantalloube\inst{\ref{ipag}} \and
A.~Cheetham\inst{\ref{geneva}} \and
P.~Delorme\inst{\ref{ipag}} \and
C.~Dominik\inst{\ref{amst}} \and
M.~Feldt\inst{\ref{mpia}} \and
R.~Galicher\inst{\ref{lira}} \and
R.~Gratton\inst{\ref{pd}} \and
J.~Hagelberg\inst{\ref{geneva}} \and
Th.~Henning\inst{\ref{mpia}} \and
M.~Janson\inst{\ref{stock},\ref{mpia}} \and
E.~Lagadec\inst{\ref{oca}} \and
A.-M.~Lagrange\inst{\ref{lira},\ref{ipag}} \and
M.~Langlois\inst{\ref{cral}} \and
C.~Lazzoni\inst{\ref{pd}} \and
H.~Le~Coroller\inst{\ref{lam}} \and
R.~Ligi\inst{\ref{oca}} \and
A.-L.~Maire\inst{\ref{ipag}} \and
G.-D.~Marleau\inst{\ref{duisburg},\ref{bernWP},\ref{mpia}} \and
F.~Ménard\inst{\ref{ipag}} \and
D.~Mesa\inst{\ref{pd}} \and
N.~Meunier\inst{\ref{ipag}} \and  
M.~Meyer\inst{\ref{umich},\ref{eth}} \and 
C.~Mordasini\inst{\ref{bernWP},\ref{bernCSH}} \and 
C.~Moutou\inst{\ref{irap},\ref{lam}} \and
A.~Müller\inst{\ref{mpia}} \and
C.~Perrot\inst{\ref{lira}} \and
M.~Samland\inst{\ref{mpia}} \and
H.~M.~Schmid\inst{\ref{eth}} \and
T.~Schmidt\inst{\ref{lira}} \and
E.~Sissa\inst{\ref{pd}} \and
M.~Turatto\inst{\ref{pd}} \and
S.~Udry\inst{\ref{geneva}} \and
A.~Zurlo\inst{\ref{diegoportales1},\ref{diegoportales2}} \and
L.~Abe\inst{\ref{oca}} \and
J.~Antichi\inst{\ref{pd}} \and
A.~Baruffolo\inst{\ref{pd}} \and
P.~Baudoz\inst{\ref{lira}} \and
J.~Baudrand\inst{\ref{lira}} \and
A.~Bazzon\inst{\ref{eth}} \and
P.~Blanchard\inst{\ref{lam}} \and
A.~J.~Bohn\inst{\ref{leiden}} \and
M.~Carbillet\inst{\ref{oca}} \and
M.~Carle\inst{\ref{lam}} \and
E.~Cascone\inst{\ref{pd}} \and
J.~Charton\inst{\ref{ipag}} \and
R.~Claudi\inst{\ref{pd}} \and
A.~Costille\inst{\ref{lam}} \and
V.~De~Caprio\inst{\ref{na}} \and
A.~Delboulb\'e\inst{\ref{ipag}} \and
K.~Dohlen\inst{\ref{lam}} \and
N.~Engler\inst{\ref{eth}} \and
D.~Fantinel\inst{\ref{pd}} \and
P.~Feautrier\inst{\ref{ipag}} \and
T.~Fusco\inst{\ref{onera},\ref{lam}} \and
P.~Gigan\inst{\ref{lira}} \and
J.~H.~Girard\inst{\ref{stsci},\ref{ipag}} \and
E.~Giro\inst{\ref{pd}} \and
D.~Gisler\inst{\ref{eth}} \and
L.~Glück\inst{\ref{ipag}} \and
C.~Gry\inst{\ref{lam}} \and
N.~Hubin\inst{\ref{eso_garching}} \and
E.~Hugot\inst{\ref{lam}} \and
M.~Jaquet\inst{\ref{lam}} \and
M.~Kasper\inst{\ref{eso_garching},\ref{ipag}} \and
D.~Le~Mignant\inst{\ref{lam}} \and
M.~Llored\inst{\ref{lam}} \and
F.~Madec\inst{\ref{lam}} \and
Y.~Magnard\inst{\ref{ipag}} \and
P.~Martínez\inst{\ref{oca}} \and
D.~Maurel\inst{\ref{ipag}} \and
O.~M\"oller-Nilsson\inst{\ref{mpia}} \and
D.~Mouillet\inst{\ref{ipag}} \and
T.~Moulin\inst{\ref{ipag}} \and
A.~Origné\inst{\ref{lam}} \and
A.~Pavlov\inst{\ref{mpia}} \and
D.~Perret\inst{\ref{lira}} \and
C.~Petit\inst{\ref{onera}} \and
J.~Pragt\inst{\ref{ipag}} \and
P.~Puget\inst{\ref{ipag}} \and
P.~Rabou\inst{\ref{ipag}} \and
J.~Ramos\inst{\ref{ipag}} \and
F.~Rigal\inst{\ref{ipag}} \and
S.~Rochat\inst{\ref{ipag}} \and
R.~Roelfsema\inst{\ref{nova}} \and
G.~Rousset\inst{\ref{lira}} \and
A.~Roux\inst{\ref{ipag}} \and
B.~Salasnich\inst{\ref{pd}} \and
J.-F.~Sauvage\inst{\ref{onera},\ref{lam}} \and
A.~Sevin\inst{\ref{lira}} \and
C.~Soenke\inst{\ref{eso_garching}} \and
E.~Stadler\inst{\ref{ipag}} \and
M.~Suárez\inst{\ref{eso_garching}} \and
Z.~Wahhaj\inst{\ref{eso_chile},\ref{lam}} \and
L.~Weber\inst{\ref{geneva}} \and
F.~Wildi\inst{\ref{geneva}}
}

\institute{
INAF-Osservatorio Astronomico di Padova, Vicolo dell'Osservatorio 5,  35122 Padova, Italy \label{pd}
\and
Department of Physics and Astronomy, University of Exeter, Stocker Road, Exeter EX4 4QL, UK \label{exeter}\email{v.squicciarini@exeter.ac.uk}
\and
Max-Planck-Institut für Astronomie, Königstuhl 17, 69117 Heidelberg, Germany \label{mpia}
\and
Universit\'e C\^ote d’Azur, Observatoire de la C\^ote d'Azur, CNRS, Laboratoire Lagrange, Nice, France \label{oca}
\and
LIRA, Observatoire de Paris, Université PSL, Sorbonne Université, Université Paris Cité, CY Cergy Paris Université, CNRS, 92190 Meudon, France \label{lira}
\and
Dipartimento di Fisica, Università di Roma Tor Vergata, via della Ricerca Scientifica 1, 00133 Rome, Italy \label{roma}
\and
Fulbright Visiting Research Scholar, Department of Astronomy, The University of Texas at Austin, 2515 Speedway, Austin, TX 78712, USA \label{austin}
\and
Institute for Astronomy, University of Edinburgh, EH9 3HJ, Edinburgh, UK \label{edinb}
\and
Aix Marseille Université, CNRS, CNES, LAM, Marseille, France \label{lam}
\and
INAF - Osservatorio Astrofisico di Catania, Via S.~Sofia 78, 95123, Catania, Italy \label{ct}
\and
Université Grenoble Alpes, CNRS, Observatoire des Sciences de l’Univers de Grenoble (OSUG), Grenoble, France \label{osug}
\and
Université Grenoble Alpes, CNRS, Institut de Plan\'etologie et
d’Astrophysique de Grenoble (IPAG), 38000 Grenoble, France
\label{ipag}
\and
Observatoire de Genève, Université de Genève, Chemin des Mailettes 51, 1290 Versoix, Switzerland \label{geneva}
\and
Anton Pannekoek Instituut, Science Park 9, 1098 XH Amsterdam, The Netherlands \label{amst}
\and 
Institutionen för astronomi, Stockholms universitet, 10691 Stockholm, Sweden \label{stock}
\and 
CRAL, CNRS, Universit\'e Lyon 1, Universit\'e de Lyon, ENS, 9 avenue Charles Andre, 69561 Saint Genis Laval, France \label{cral}
\and
Fakult\"at f\"ur Physik, Universit\"at Duisburg--Essen, Lotharstra\ss{}e 1, 47057 Duisburg, Germany \label{duisburg}
\and
Division of Space Research \&\ Planetary Sciences, Physics Institute, University of Bern, Gesellschaftsstr.~6, 3012 Bern, Switzerland \label{bernWP}
\and
Center for Space and Habitability, University of Bern, 3012 Bern, Switzerland \label{bernCSH}
\and 
Department of Astronomy, University of Michigan, Ann Arbor, MI 48109, USA \label{umich}
\and
Institute for Particle Physics and Astrophysics, ETH Zurich, Wolfgang-Pauli-Strasse 27, 8093 Zurich, Switzerland \label{eth}
\and
Université de Toulouse, CNRS, IRAP, 14 avenue Belin, 31400 Toulouse, France \label{irap}
\and
Instituto de Estudios Astrof\'isicos, Facultad de Ingenier\'ia y Ciencias, Universidad Diego Portales, Av.~Ej\'ercito Libertador 441, Santiago, Chile \label{diegoportales1}
\and
Millennium Nucleus on Young Exoplanets and their Moons (YEMS), Santiago, Chile \label{diegoportales2}
\and
Sterrewacht Leiden, Universiteit Leiden, PO Box 9513, 2300 RA Leiden, The Netherlands \label{leiden}
\and 
INAF - Osservatorio Astronomico di Capodimonte, Salita Moiariello 16, 80131 Napoli, Italy \label{na}
\and 
ONERA (Office National d'Études et de Recherches Aerospatiales), B.P.72, 92322 Chatillon, France \label{onera}
\and 
Space Telescope Science Institute, 3700 San Martin Drive, Baltimore, MD, 21218, USA \label{stsci}
\and
European Southern Observatory (ESO), Karl-Schwarzschild-Str.~2, 85748 Garching, Germany \label{eso_garching}
\and
NOVA Optical Infrared Instrumentation Group, Oude Hoogeveensedijk 4, 7991 PD Dwingeloo, The Netherlands \label{nova}
\and
European Southern Observatory, Alonso de Córdova 3107, Vitacura, Casilla 19001, Santiago, Chile \label{eso_chile}
}

\date{Received 15 January 2026 / Accepted --}
 
\abstract
{Unbiased surveys of large stellar samples are the prime means through which the prevalence of exoplanets can be derived, and crucial constraints to planet formation models can be set. Direct imaging (DI) is  ideally positioned to probe the outer regions (5--300\,au) of planetary systems, providing complementary information to techniques such as transits and radial velocities.}
{We present the full sample of the SpHere INfrared survey for Exoplanets (SHINE), the second largest DI campaign to date. SHINE observed 460 stars between 2015 and 2023 thanks to the guaranteed time observations (GTO) allocated by ESO to the SPHERE consortium at VLT. The goal of this paper is to homogeneously derive the stellar properties of the targets and to define a subsample of young single hosts to be used as a starting point for the final statistical analysis of the survey.}
{Stellar ages were determined based on kinematic indicators (such as the membership to young moving groups), age diagnostics (lithium abundance, rotation,  and activity), and isochrone fitting. A thorough vetting for binarity was undertaken combining astrometric, spectroscopic, and imaging data.}
{A subsample of 333 stars, covering a large extent of stellar ages and masses, was constructed. Selection criteria, global features, as well as the properties of individual stars are reported and discussed.}
{}

\keywords{stars: fundamental parameters
 -- stars: pre-main sequence
 -- (stars:) binaries: general
 -- planets and satellites: detection
 }

\titlerunning{The SPHERE infrared survey for exoplanets (SHINE). V.}
\authorrunning{V. Squicciarini, S. Desidera et al.}

\maketitle
\nolinenumbers

\section{Introduction}
\label{sec:intro}

Over the past three decades, exoplanetary studies have transitioned from a pure detection focus to the pursuit of probing the physical diversity, architectures, and origins of planetary systems. With more than 6000 confirmed exoplanets to date, the attention has progressively shifted toward the statistical interpretation of global system properties -- what is often called exoplanet demographics \citep[e.g.,][]{winn15}. By tracing how physical and orbital properties depend on stellar and environmental factors, demographic studies provide a framework to test models of planet formation and evolution \citep[e.g.,][]{gaudi21,biazzo22}.

As no single detection technique can cover the full space of planetary parameters, a comprehensive picture requires the combination of complementary methods,  with each being sensitive to different regions in mass, age, and semi-major axis. Among these, radial velocity (RV) and direct imaging (DI) offer particularly synergistic views to study the prevalence and the properties of giant planets ($M \gtrsim 0.3~M_{\text{Jup}}$): while RV surveys have mapped the old population within $\sim$5\,au \citep{wolthoff22,rosenthal24}, DI extends sensitivity to younger  ($t_* \lesssim 1~\text{Gyr}$), more separated objects beyond roughly 10--20\,au. This is particularly significant since RV results become less reliable when extended to the long-period regime \citep{lagrange23}, resulting in extrapolated yields being too optimistic \citep{dulz20}. DI searches probe a separation domain where different formation channels may operate -- core accretion \citep{pollack96,mordasini09}, which is expected to be the main formation channel of RV planets, and its pebble accretion modification \citep{ormel10,johansen21}; and gravitational instability \citep{boss97,kratter16}. Beyond discovery, DI uniquely allows the direct spectrophotometric characterization of young giant planets, providing access to information about the temperature, composition, cloud properties, and atmospheric dynamics, and therefore placing constraints on atmospheric chemistry, thermal evolution, and formation history \citep{currie23}.

Since the first DI detection in 2004 \citep{chauvin04}, roughly 50 planetary ($M < 13~\Mjup$) companions have been identified\footnote{Based on the Extrasolar Planet Encyclopaedia: \url{https://exoplanet.eu/catalog/}.}, comprising benchmark systems such as the accreting PDS~70 b and c \citep{keppler18,muller18,mesa19}, the $\sim$20~Myr-old 51~Eridani b \citep{macintosh15}, $\beta$~Pictoris b \citep{lagrange09}, and AF~Leporis b \citep{mesa23,derosa23,franson23}, as well as the four planets in the HR~8799 system \citep{marois08}. More recently, JWST has extended DI to colder and more mature giant planets \citep{matthews24,lagrange25b,bardalez25}. 
Most of these discoveries stem from large-scale, blind surveys that have observed samples ranging from a few tens of stars -- such as MASSIVE \citep{lannier16}, SEEDS \citep{uyama17}, LEECH \citep{stone18}, and BEAST \citep{janson21} -- to several hundreds, as in the NICI-PCF \citep{liu10}, IDPS \citep{galicher16}, and ISPY-NACO \citep{launhardt20} programs. The current state of the art in DI surveys is represented by the SpHere INfrared survey for Exoplanets \citep[SHINE;][]{chauvin17} and the Gemini Planet Imager Exoplanet Survey \citep[GPIES;][]{nielsen19}, which leveraged instruments aboard adaptive-optics-fed 8-m-class telescopes to
explore samples of approximately 400 and 600 stars, respectively. {{red}F}or both surveys, however, only partial statistical analyses based on $\sim 50\%$ of stars have been published \citep{nielsen19, vigan21}.

Starting in 2015, the SHINE survey carried out its observations until 2023 during 200 nights allocated by ESO as guaranteed time observations in exchange for the construction of the Spectro-Polarimetric High-Contrast Exoplanet Research (SPHERE) instrument \citep{beuzit19}. The scientific goals of SHINE include the discovery of new substellar companions amenable to follow-up characterization studies; the link between planets and disks; the study of global architectures of multi-object systems; the determination of the occurrence frequency of giant planets and brown dwarfs in the [5, 300]\,au range; the comparison of these constraints with formation models; and the dependence of these trends on stellar mass and age. 

While SHINE was $30\%$ complete, a series of three papers presented the early results emerging from the campaign \citep{desidera21,langlois21,vigan21}. The first publication (hereafter \citetalias{desidera21}) introduced the so-called F150 sample, defined as the set of stars with a first epoch obtained before February 2017, after a general overview of the target selection process for the entire survey. The second paper focused on the observations and the data reduction, while the third paper presented the statistical analysis of the F150 sample. With the survey that is now complete, we are ready to present the results emerging from the entire sample (hereafter F400). Following the same scheme designed for the intermediate analysis, we split the description of the survey into three papers: a first one (this paper) presenting an overview of the stellar sample with its main astrophysical properties; a second publication \citep{chomez25} detailing the performances attained during the campaign, both in terms of raw instrumental contrasts and after post-processing; and finally, a third paper \citepalias{chauvin26} dealing with the extraction of statistical constraints from F400 and a comparison with planet formation models.

This paper is organized as follows: Section~\ref{sec:shine_intro} summarizes the target selection and observation criteria, highlighting the updates compared to \citetalias{desidera21}. 
Section~\ref{sec:params} focuses on the main goal of this paper, namely, the derivation of stellar parameters for the entire stellar sample. 
The statistical sample is defined and discussed in Section~\ref{sec:statistical_sample}. In Section~\ref{sec:discussion} we present a general overview of the properties of the statistical sample and empirically assess the level of residual binary contamination. 
Finally, Section~\ref{sec:conclusion} sums up the main results of this paper.
Extensive notes on how stellar properties were determined for individual stars are provided in the appendices. In particular, Appendix~\ref{a:notes_f150} describes the targets that were already part of F150, while Appendix~\ref{a:notes_f400} focuses on the newly added stars; finally, Appendix~\ref{a:notes_removed} explains the reasons for excluding specific targets from the statistical sample.

\section{The SHINE survey}
\label{sec:shine_intro}

\subsection{Survey design}
\label{sec:design}

The design of SHINE and the various stages of selection of the stellar sample were described in detail in \citetalias{desidera21}, and only briefly recapped here for convenience. As a first step, some general selection criteria, mostly based on magnitude, distance, mass, and age, were used to design a large sample of 1224 potential targets. In particular,  an upper stellar mass limit at $3~M_\odot$ was imposed\footnote{The B-Star Exoplanet Abundance STudy \citep[BEAST;][]{janson21,delorme24} was subsequently designed to complement SHINE in the B-type regime.}; known spectroscopic binaries and systems with at least one stellar companion within $6"$ (i.e., within the SPHERE field of view) were excluded to both maximize sensitivity to planetary-mass companions and maintain a homogeneous sample suitable for studying planet formation around single stars. Although this initial list was defined in 2014, newly identified binaries were removed as the survey progressed.

The initial star list was split into priority groups according to a figure of merit determined from  the stellar properties (mass, age, distance, magnitude) and the expected detectability of planets considering power-law distributions for planetary mass and semi-major axis following \cite{cumming08}. Several simulations were performed, considering two different values of the cut-off in semi-major axis (15 and 30\,au), and assuming a scaling or a lack of scaling of the planetary mass with stellar mass.
From a combination of the outcome of these simulations (to avoid a sample optimized on a specific distribution of planetary companions), 800 stars were selected and evenly distributed in  four priority bins. This roster includes stars belonging to the Scorpius-Centaurus association (Sco-Cen) and to other young moving groups and associations (YMGs), as well as young stars ($t_\star<1$\,Gyr) belonging to the field. 
 A fixed number of 40 Sco-Cen stars was considered for each priority bin for operational constraints (right ascension distribution and expected number of field objects at the low galactic latitude of the association).
A fifth priority bin was added, comprising bright stars to be targeted with short observations during  nonoptimal weather conditions or short schedule gaps (Table~\ref{tab:priority}).
 These included young close binaries observed for orbit monitoring \citep{calissendorff2022} and bright stars with known planets discovered with the RV technique.
Finally, a category of targets with special priority, labeled as P0, was added, including stars with special motivations for observations with SPHERE, such as known substellar companions from DI or resolved circumstellar disks. The P0 stars that were not already part of the 800-star sample will not be considered in this work and in the subsequent statistical analysis.

Starting from ESO P98 (April-September 2016), the sample was complemented with $\sim 50$ M-type members of YMGs following a re-evaluation of the performances of SPHERE at the faint ($R=12$  mag) end. These stars were assigned to priority bins P1 and P2 \citepalias{desidera21}. In addition to this, a reassessment of the priorities of Sco-Cen targets happened halfway through the survey. In particular, during the 2017 season the original priorities were modified by one unity to increase the probability of Sco-Cen stars to be observed: P1 $\rightarrow$ P0, P2 $\rightarrow$ P1, P3 $\rightarrow$ P2, P4 $\rightarrow$ P3. From 2018, all Sco-Cen stars with no first epoch were flagged as P0 or P1 depending on the galactic latitude ($|b|>7^\circ$ and $|b|<7^\circ$, respectively), to give higher priority to the targets expected to have less contamination from background interlopers.
In the last two years of the survey (ESO periods P104-P107, Prog. ID 1104.C-0416), second-epoch observations were given higher priority, depending on the ranking of the candidates. 
Few additional first-epoch observations were obtained during nights when no targets with prior observations were available for follow-up.

The computation of the statistical weights for the different bins and the impact of the observations of targets with special priority in the statistical analysis with be detailed in a forthcoming paper \citep{chauvin26}. We anticipate, for the sake of clarity, that Sco-Cen and non Sco-Cen targets were treated independently so as to account for the bias induced by the mid-survey reassessment of Sco-Cen priorities.

Besides second-epoch observations during the GTO program, a dedicated open time program (ESO Prog. ID 0110.C-4198, 0111.C-0196, 0113.C-2177) named snapSHINE was carried out to maximize the amount of follow-up observations of the candidates detected during the first epochs. During this snapshot survey, only follow-up observations were performed. We defer the reader to \citet{chomez25} for a detailed discussion about the observations, the snapSHINE survey, and the data analysis.

\begin{table}[t]
\caption{Priority distribution of the SHINE sample.}
\centering
\begin{tabular}{lcc}
\hline \hline
Bin                & Early-type stars  & Solar and Low-mass stars \\
\hline
P0                      &  \multicolumn{2}{c}{Special targets}  \\
P1                      & 20 YMGs + 40 Sco-Cen & 120 YMGs + 20 Field \\
P2                      & 20 Field + 40 Sco-Cen & 50 YMGs + 90 Field \\
P3                      & 20 Field + 40 Sco-Cen & 140 Field \\
P4                      & 20 Field + 40 Sco-Cen & 140 Field \\
P5                      &  \multicolumn{2}{c}{Bad weather backup or filler} \\
\hline
\end{tabular}
\label{tab:priority}
\end{table}

\subsection{Observation setup}
\label{sec:obs_setup}

The survey was mostly operated in the \textsc{IRDIFS} mode, that allows for simultaneous spectroscopic observations in the $YJ$ range
(0.95--1.35$~\upmu$m, $R \sim 50$) over a small
field of view ($1.77" \times 1.77"$) by means of the Integral Field Spectrograph \citep[IFS;][]{claudi08} and dual-band photometric observations in the $H$ band ($1.593~\upmu$m, $1.667~\upmu$m) over a larger $11" \times 11"$ field of view with the infra-red dual imaging and spectrograph \citep[IRDIS;][]{dohlen08}.
For Sco-Cen targets, an alternative configuration named \textsc{IRDIFS-EXT} was used
starting from March 2017,
where IFS spans $YJH$ bands and IRDIS is equipped with two narrow filters in the $K$ band ($2.110~\upmu$m, $2.251~\upmu$m).
The \textsc{IRDIFS} setup was employed in snapSHINE; in this case, we decided to collect short exposures (typically $\sim 30$ min, including overheads) to follow up as many companion candidates as possible.

\subsection{The F400 sample}
\label{sec:sample}

The definition of the priority bins and the large sample size were devised to optimize the scheduling of the observations; indeed, based on the number of available nights ($\sim 200$), we expected only about 400 stars to be observed. At the end of the survey, 460 targets had at least one epoch of acceptable quality; we consider this sample (hereafter F400 sample) as the starting point for our analysis.

The main astrometric, kinematic, photometric and spectral parameters for the sample are provided in Tables D.1, D.2, and D.3.
Equatorial coordinates, $G$ magnitudes and radial velocities were collected from Gaia DR3 \citep{gaiadr3}; as regards proper motion, we favored whenever available the long-term proper motion computed as the difference between Gaia DR3 and {\it Hipparcos} \citep{vl07} coordinates divided by their epoch difference (24.75 yr), so as to minimize the perturbation from short-period binaries when assessing the membership to YMGs (Section~\ref{sec:kin}). Spectral types were collected from Simbad \citep{simbad}, while $H$ magnitudes come from 2MASS \citep{skrutskie06}. The presence of disks (Section~\ref{sec:disks_and_inner_planets}), either resolved or inferred from an infrared excess, is also reported. 

\section{Updated stellar properties}
\label{sec:params}

The determination of stellar ages is a particularly delicate step for the correct interpretation of direct imaging surveys, as the thermal emission of self-luminous giant planets and brown dwarfs decreases over time. Likewise, stellar masses are a crucial parameter for a campaign such as SHINE that has as one of its main goal to explore the dependence of the properties of wide-orbit planets on host mass.

 Isochrone fitting is one of the most widely used methods for estimating stellar ages and other fundamental stellar parameters. For the purpose of this work, we employed {\tt{MADYS}} \citep{squicciarini22} -- a versatile tool allowing for parameter determination for stellar and substellar objects based on the comparison between the available photometric information and evolutionary models. We used the PARSEC isochrones \citep[v2.0;][]{nguyen22} due to their large dynamical range covering the entire age and mass range of our interest. Extinction was typically computed by integration of the 3D extinction map by \citet{leike20} -- with a few exceptions detailed in the Appendices.

The  ages and masses determined in this work are provided in Table D.4. 
We describe in detail the age determination procedure in Section~\ref{sec:stellar_ages}, after presenting the different age indicators employed in this work in the following Section~\ref{sec:age_indicators}.

\subsection{Age indicators}
\label{sec:age_indicators}

\subsubsection{Kinematics and moving group membership}
\label{sec:kin}

As in \citetalias{desidera21}, we collected the most up-to-date astrometric and kinematic information for our targets to assess their membership to young moving groups and associations. Equatorial coordinates ($\alpha$, $\delta$), parallaxes ($\varpi$), and radial velocities (RV) were taken from Gaia DR3, which provides improved precision and accuracy compared to the earlier Gaia DR2 \citep{gaia_dr2} available at the time of the F150; as regards proper motions ($\mu_\alpha \cos{\delta}$, $\mu_\delta$), as mentioned in Section~\ref{sec:sample}, we employed long-term {\it Gaia-Hipparcos} measurements instead of Gaia DR3 proper motions. Given the longer timespan (24 yr) compared to Gaia DR3 (2.8 yr), these values are expected to be less affected by astrometric perturbations when dealing with binary systems with an orbital period of a few years \citep{kervella19}, which are generally not resolved in our SPHERE observations. 

The evaluation of membership probabilities was performed using the \banyan~ tool\footnote{\url{http://www.exoplanetes.umontreal.ca/banyan/banyansigma.php}.} \citep{banyansigma} based on the six parameters ($\alpha$, $\delta$, $\varpi$, $\mu_\alpha \cos{\delta}$, $\mu_\delta$, RV). The membership probabilities to a YMG  ($p_{\text{YMG}}$) or to the field are in 90\% of the cases larger than 0.8 and in 85\% of the cases larger than 0.9. For stars with $p_{\text{YMG}} < 0.8$, the reliance on indirect indicators (activity, lithium, rotation) was crucial to confirm or reject membership.

The results of this analysis are reported in Table D.2.
Each YMG was assigned an optimal, minimum and maximum age ($t^*_\text{opt}$, $t^*_\text{min}$, $t^*_\text{max}$) based on the literature studies reported in Table~\ref{t:mgage}. We point out that the TW Hya moving group was split in two subgroups based on the recent analysis by \citet{miretroig25}.

{ 
\begin{table}[t]
\tiny
\caption{Adopted YMG ages.}
\centering
\begin{tabular}{lcccc}
\hline \hline
   Group (acronym)       & $t^*_\text{opt}$ &  $t^*_\text{min}$  & $t^*_\text{max}$  & Ref. \\
                            & (Myr) & (Myr) & (Myr) & \\
\hline
AB Doradus (ABDO)           & 137 & 120 & 154 &  1\\
Argus (ARG)                 &  48 &  38 &  58 &  1 \\
$\beta$ Pic (BPIC)          &  21 &  17 &  25 & 1 \\
Carina (CAR)                &  28 &  17 &  39 &  1 \\
Carina-Near (CARN)         &  200 & 180 & 220 &  1 \\
Columba (COL)               &  36 &  28 &  44 & 1 \\
$\varepsilon$ Cha (EPSC)    &  4 &  2 &  8 & 2 \\
$\eta$ Cha (ETAC)        &  11 &   8 &  14 & 3  \\
Lower Centaurus-Crux (LCC)  &  16 &  12 &  20 & 3 \\
Tuc-Hor (THA)               &  37 &  26 &  48 & 1 \\
TW Hya (TWA) a                & 9 &  8 &  11 & 4\\
TW Hya (TWA) b               &  6 &  5 &  8 & 4\\
Upper Centaurus-Lupus (UCL) &  17 &  15 &  20 & 3 \\
Upper CrA (UCRA)         &  15 &  13 & 17 & 5 \\
Upper Scorpius (US)         &  11 &   4 &  12 & 3 \\
Ursa Major (UMA)         &  414 & 391 & 437 & 6 \\
\hline
\end{tabular}
\label{t:mgage}
\tablefoot{References: 1: \citet{gratton24}; 2: \citet{murphy13}; 3: \citet{desidera21}; 4: \citet{miretroig25}, 5: \citet{esplin22}; 6: \citet{jones15}.}
\end{table}
}

\subsubsection{Wide stellar companions and comoving stars}
\label{sec:cms}

As mentioned in Section~\ref{sec:design}, the SHINE sample was vetted for short- and intermediate-separation binaries ($\rho < 6"$); no cut was operated with respect to stars having wider companions. Gaia DR3 is virtually complete for stellar objects at the distances probed by SHINE, and can be reliably used to check for physical associations of stars by means of parallax and common proper motion. In this way, we could identify 26 stars (Table D.5) hosting 30 wide stellar companions; in two cases, namely the Fomalhaut triple system (HIP 113283, HIP 113368, LP 876-10) and the HD 61606 binary system (HIP 37349, HD 61606B), all stellar components were targeted by the survey.

Under the assumptions that the components of wide binaries are coeval, the secondaries resolved in {\it Gaia} could be used as additional data points for the isochronal analysis; this sometimes allowed us to break the degeneracies in the (mass, age) space. 
For each star belonging to Sco-Cen, following the method designed by \citet{janson21}, we looked for kinematic neighbors in {\it Gaia}, that we define comoving stars (CMSs). Typically composed by a few tens to a few hundreds of objects, CMS groups -- which are not gravitationally bound and are bound to disappear on timescales of a few $10^7$ yr -- result from common star formation events and capture the complex age variability within the region in a more accurate way than the standard subgroup classification (although residual, smaller-scale age structure might still exist and impact the resulting estimates). We identified 82 CMS groups; for each of these, the age and the age uncertainty are determined from a weighted mean and variance of the isochronal ages of its components, and assumed to be an age indicator for the corresponding SHINE star in addition to its individual isochronal age.

\subsubsection{Lithium}

The equivalent width (EW) of the \ion{Li}{i} $\lambda6707.8$\,{\AA} doublet is a powerful age diagnostic for late-type objects.
The available measurements were compiled from the literature.
For additional 12 stars in our sample without determinations or with discrepant measurements in the literature, the EW(Li) was measured on publicly available high-resolution spectra
from FEROS \citep{kaufer98}, HARPS \citep{pepe02}, SOPHIE \citep{perruchot08}, and CHIRON \citep{chiron13} spectrographs\footnote{ESO archive for HARPS and FEROS reduced spectra: \url{https://archive.eso.org/cms.html}; SOPHIE archive: \url{http://atlas.obs-hp.fr/sophie/}; NOIRLab archive for CHIRON: \url{https://astroarchive.noirlab.edu/portal/search/}.}.
The archival spectra were continuum normalized using synthetic spectra generated with {\tt MOOG} (\citealt{sneden1973} -- 2019 version), adopting appropriate stellar parameters for our stars using {\tt iSpec} \citep{blanco-cuaresma}. The same template spectrum was applied to each observed spectrum to obtain the radial velocity corrections to the rest frame.
Additionally, we used the {\tt ARES v2} code developed by \cite{sousa} to determine the equivalent widths of the \ion{Li}{i} doublet, adjusting the fitting parameters (such as signal-to-noise ratio and full width at half maximum) as necessary for the various data sets. Errors were calculated by considering the continuum placements and potential line blending. For more details, we refer the reader to \cite{sousa}. These new measurements are listed in Table D.3 and included in the derivation of the system age.
 The age is inferred through a comparison of empirical sequences of EW(Li) vs.photometric colors of clusters or groups of well known age, as in \cite{desidera15}. Updated empirical sequences were adopted when available \citep[e.g.][for BPIC]{messina16}. The calibration is qualitatively similar although fully independent to the one presented by \cite{jeffries2023}.

\subsubsection{Photometric variability and rotation period}

Gyrochronology is becoming increasingly popular as a method for age determination, thanks to the availability of high-quality photometric light curves from the ground \citep[e.g. ASAS survey;][]{pojmanski02} and especially from space, thanks to the missions searching for transiting planets as TESS \citep{ricker15} and {\it Kepler 2} \citep[K2;][]{howell14}. 
Rotation periods for our target stars were compiled considering both kinds of datasets; no new period was determined in this work. \citet{messina10}, \citet{messina11}, \citet{kiraga12}, and \citet{desidera15} were the main sources for ground-based observations, \citet{desidera21}, \citet{tu20}, and \citet{fetherholf23} for space-based ones.
The observed rotation period emerging from the observations is in some cases a harmonic of the true period.  Checks with projected rotational velocity
$ v \sin i$ and with the expected rotation period
 of chromospheric emission $\log R'_{HK}$ \citep{mamajek08} were performed.
Details for individual sources are provided in Appendices~\ref{a:notes_f150}-\ref{a:notes_removed}.

\subsubsection{Chromospheric and coronal activity}

The Ca~{\sc ii} H\&K index for chromospheric emission, $\log R'_{HK}$, was compiled from the literature. Most of the sources adopted the prescriptions of the M.~Wilson survey \citep{Wilson1968,baliunas95} for the determination of the chromospheric emission. Higher weight was given to sources based on multi-epoch measurements in order to consider the variability on the timescales of rotational modulations and activity cycles. 
Coronal emission was obtained from the ROSAT All Sky Catalogs \citep{rosatbright,rosatfaint}. The X-ray fluxes were obtained using the calibration by \citet{hunsch99}, including the hardness ratio HR1  and using the adopted distance to obtain the X-ray luminosities; a cross-match radius of 30" was adopted.

The use of these indicators as age diagnostics assumes a well-defined relationship
between activity and age, as resulting, for example, from the standard age calibrations by \cite{mamajek08}. This has been recently questioned by the increasing evidences of spin-down stalling \citep[see ][ and references therein]{santos2025} but this feature appears to be relevant only or mostly at ages older than about 1\,Gyr, therefore outside the boundaries of our sample.
The dependence of chromospheric and coronal activity on age becomes limited in the
so-called saturation regime \citep[e.g., ][]{pizzolato2003,wright11}, corresponding to ages younger than $\sim$ 100\,Myr for G and K stars. Furthermore, the age dependence becomes weaker over the whole range of ages for mid-early F-type stars, due to vanishing of the outer convective envelope \citep{desidera15}.
The role of metallicity in the age calibration \citep{carvalhosilva2025} would have minor impact for our target stars, considering the general evidence for chemical composition close to solar for nearby young stars.

\subsection{Age determination}
\label{sec:stellar_ages}

The procedure followed to converge toward a final age estimate varied depending on the available data (Table D.3),
on the spectral type (Table D.1),
and on the membership status to YMGs (Table D.2).
The decision tree for the determination of stellar age is based on
the following criteria: 

    {\it Members of YMGs}. For targets that result high-probability members of YMGs from
    kinematic analysis, we verified the consistence of the known YMG age with the age provided by isochrones and, whenever available, to indirect indicators. In the vast majority of cases this consistency check is positive and the age of the corresponding YMG from Table \ref{t:mgage} is adopted. 
    Targets with ambiguous membership or discrepant indicators are evaluated case by case, typically adopting the YMG age but extending the errorbars encompassing the results of other methods when group membership is considered the most probable outcome, or adopting the age from other methods and including the YMG age in the errorbars when the membership is ambiguous.
    
    {\it Sco-Cen stars}. The logic is similar to the previous case but we aim at
    a finer age determination than the age of the three main regions of Sco-Cen, considering the robust evidence of distinct subgroups among the various regions.
    Three indicators, all based on isochrones, were derived in this case. Ordering them in ascending order of spatial scale, we distinguish the individual isochrone fitting, the group fitting of the CMSs, and, on the largest scale, the subgroup age as from Table~\ref{t:mgage}. In the majority of cases, the indicators are consistent with each other. Individual age measurements that are distinct from the other two indicators often stem from unresolved binarity, but sometimes reflect real features of the systems.
    For stars with ambiguous membership, the same criteria as for the previous item were adopted.
    
    {\it Field objects}. For targets with low or null probability of membership,
    the age determination is performed considering the indicators for the individual object
    and, when applicable, for confirmed comoving companions. The priority of the individual indicators depends on the spectral type stars and on the plausible age range for the target:
    \begin{itemize}
        \item {\it Early-type (BA) field stars}. Given the short pre-main sequence phase and the rapid evolution of these objects, isochronal analysis is sufficient to unambiguously determine an age estimate with a 10-30\% uncertainty. These estimates are corroborated by indirect indicators, whenever available.
        \item {\it Early-mid F-type field stars}. These stars are those with largest age errors in most cases, as the isochrone age has typically large uncertainties while indirect indicators have poor sensitivity to stellar age. We exploit the combination of all available information to constrain the system age.
        \item {\it Late-type (FGKM) field stars}. Indirect indicators (rotation, activity, lithium) are crucial to determine the age of these stars, as the slow temporal evolution along the main sequence (MS) makes isochrone constraints very loose and often degenerate between a pre-MS and a post-MS solution. The presence of wide stellar companions turned out to be extremely useful in some cases. The procedures from \citet{desidera15} were adopted to estimate the system ages from these indicators. As general criteria, gyrochronology is considered as the most reliable indicator in the range 100-150\,Myr to 1\,Gyr (although not for M stars; see \citealt{cortes24} and references therein) while at younger ages it does not provide
        a well-defined value. Below 100\,Myr, lithium is typically more sensitive, depending on star spectral type, and isochrone pre-MS ages are often valuable. 
        Chromospheric and coronal activity are considered of lower priority, as their dependence on stellar age is linked to the rotational evolution of the star and they
        are subject to variability due to rotational modulations, flares and activity cycles.
        Nevertheless, they are considered when the determination of the rotation period is missing or ambiguous.  The \vsini~ measurements are used as a consistency check for the rotation period. 
    \end{itemize}

\begin{figure*}
    \centering
    \includegraphics[width=0.95\linewidth]{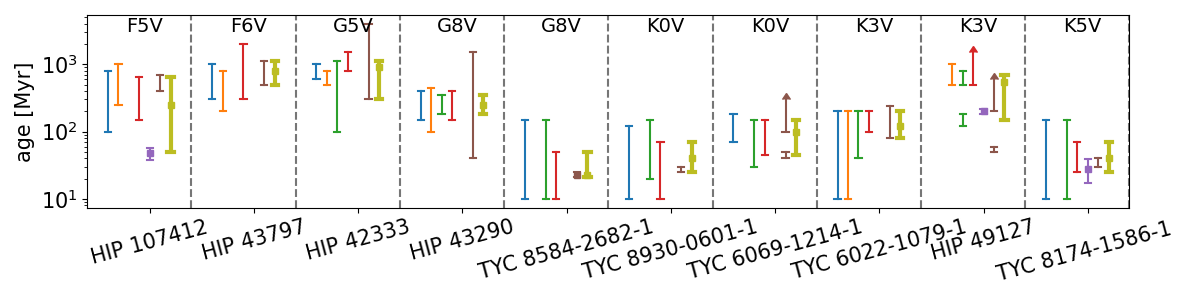}
    \caption{Age estimates for ten representative stars with no clear YMG membership. The ages obtained via different methods are shown as ranges or lower limits: X-ray activity (blue), chromospheric activity (orange), rotation (green), lithium (red), YMG (lilac), isochrones (brown). The rightmost values represent the final adopted ages.}
    \label{fig:age_examples}
\end{figure*}

\vspace{0.2cm}
\noindent The final age determinations are provided in Table D.4.
{In the majority of cases, the individual age indicators -- whose sensitivity depends, as already mentioned, on spectral type and on the absolute age value -- agree within individual errors (see Fig.~\ref{fig:age_examples}). The most numerous class of ambiguous cases refers to objects with moderate membership probability to some of the YMG and with uncertain (or marginally discrepant) results from age indicators (15 targets). In this case, the YMG age or the age from indicators were adopted depending on the confidence of the membership assignment, allowing for the alternative explanation through the error bar. 26 additional targets have minimum or maximum ages which differ by the nominal age by more than a factor of three; in the majority of the cases the large uncertainties are due to the availability of just one method (typically isochrones). Several other cases, initially flagged as discrepant, were cleared after a dedicated analysis; possible causes for the discrepancies included rotation periods being aliases of true ones or belong to other stars, unrecognized or spurious multiplicity, underestimated errors, or the presence of a tidally-locked component.
We provide in Appendices~\ref{a:notes_f150}-\ref{a:notes_removed} a case-by-case description of the results for all targets with no clear YMG memberships, unless a note existed in \citetalias{desidera21}} and was not updated since.

\subsection{Stellar masses}

Based on the ages derived in Section~\ref{sec:stellar_ages}, we employed again {\tt{MADYS}} to compute stellar masses.  The corresponding uncertainties account for the uncertainties in parallax, photometry, and age, but not for the systematics arising from the choice of a specific stellar model with a fixed solar metallicity.
For unresolved binaries (Section~\ref{sec:binaries}), we disentangled the photometry of the two components based on the mass ratio $q$ -- whether available or loosely constrained. The mass uncertainty also incorporates the uncertainty on $q$ in these cases.

\section{Statistical sample}
\label{sec:statistical_sample}

\subsection{Vetting for binarity}
\label{sec:binaries}

The SHINE survey was designed to target single stars for a twofold reason: firstly, because the presence of a bright secondary companion can deteriorate the quality of high-contrast imaging observations; secondly, because planet formation processes likely occur in a different way around multiple objects -- not to mention the enhanced possibilities of dynamical evolution compared to systems that are gravitationally dominated by a single central body. For these reasons, we seek to define a subsample of bona-fide single hosts for the purpose of the final statistical analysis \citepalias{chauvin26}.

Operatively, we distinguished three kinds of stellar companions: those too close to be resolved by SPHERE, those within the IRDIS field of view ($\sim 6"$), and those having an even larger projected angular separation. We removed five stars with a stellar companion belonging to the third category and such that its projected separation $s< 500$\,au. This somewhat arbitrary threshold was selected based on the fact that stellar multiplicity has been shown to affect giant planet occurrences and system architectures in binaries up to several hundred au separations \citep{holman99,fontanive19,moe21,fontanive21,cadman22}, restricting, for instance, the semi-major axes where stable orbits can exist \citep{holman99}. Likewise, we removed all the stars with companions belonging to the second category, most of which had been already identified and discussed in \citet{bonavita22}.

In order to identify the first category of companions, we first checked whether each star was listed in the Washington Double Star catalog \citep{mason01} or indicated as multiple in RV studies in the literature. This allowed us to identify eight spectroscopic binaries. Afterwards, we made use of the large wealth of data products provided by {\it Gaia} to search for unknown companions: even if not resolved, massive secondaries imprint wiggles on their primaries' proper motions that can be measured by the instrument. In the most conspicuous cases, stars are listed in the {\it Gaia} non-single star catalog \citep[GaiaNSS;][]{gaia_nss}, a catalog of sources showing a nonlinear proper motion compatible with a Keplerian acceleration. Less prominent companions can be identified by means of the RUWE parameter \citep{lindegren18} and through the {\it Gaia-Hipparcos} proper motion anomaly \citep[PMa;][]{kervella19,kervella22}, where the latter quantity is typically sensitive to larger periods compared to the former due to the longer timespan of {\it Gaia-Hipparcos} relative astrometry (Section~\ref{sec:kin}).

\begin{figure}
    \centering
    \includegraphics[width=0.95\linewidth]{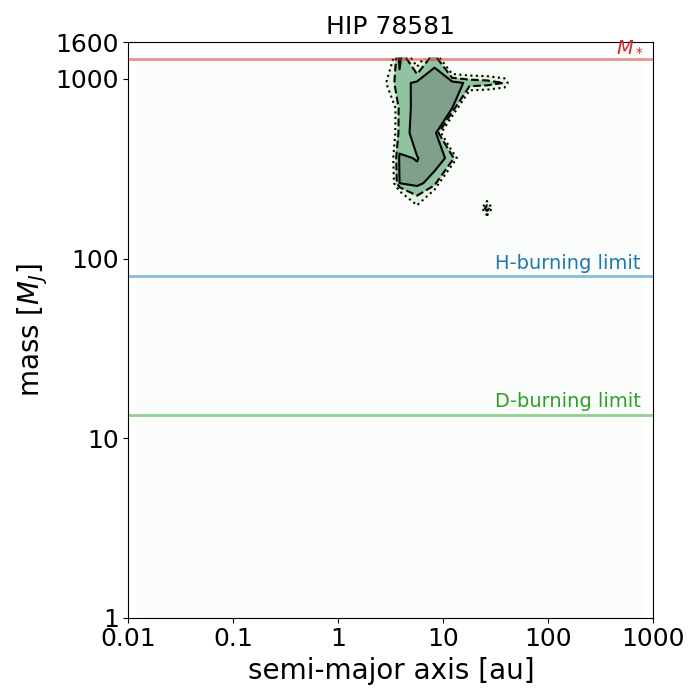}
    \includegraphics[width=0.95\linewidth]{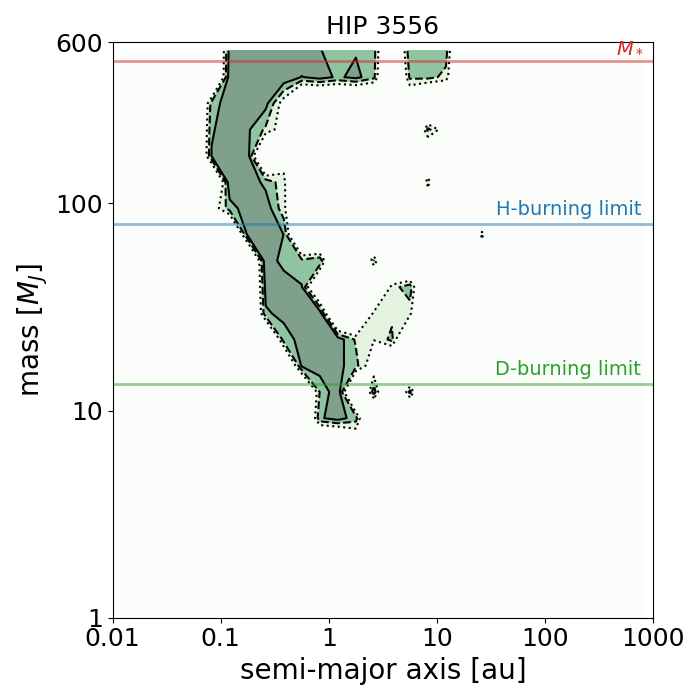}
    \caption{Constraints from {\tt GaiaPMEX} on the mass and semi-major axis of the companions responsible for the RUWE and PMa signal measured for HIP 78581 (top) and HIP 3556 (bottom). The colored regions, delimited by contours, correspond to 68.3\%, 95.4\%, and 99.73\% confidence intervals. }
    \label{fig:gaiapmex_example}
\end{figure}

It is common in the literature to use a constant threshold $\ruwe = 1.4$ to identify binary stars; however, the criterion has been shown not to be particularly accurate \citep[see, e.g.,][]{castroginard24}. To address this limitation, we employed {\tt GaiaPMEX} \citep{kiefer25}, a tool that evaluates the statistical significance of both RUWE and PMa under the null hypothesis of a single-star solution. This approach relies on a detailed modeling of the astrometric noise budget as a function of the {\it Gaia} $G$ magnitude and $BP-RP$ color, yielding approximately normal distributions for the single-star UEVA$^{1/3}$ and PMa$^{2/3}$ (where the unbiased estimator of variance a posteriori (UEVA) is a quantity related to RUWE). We denote the resulting significances of the two astrometric signatures $\alpha_{\text{UEVA}}$ and $\alpha_{\text{PMa}}$ (Eq.~27 and Eq.~29 of \citealt{kiefer25}) as $\snrruwe$ and $\snrpma$, respectively. The observed signatures can be used to set constraints on the possible mass and semi-major axis of the unseen companion. Two examples of {\tt GaiaPMEX} maps with a significant signal, one caused by a stellar-mass astrometric companion and one by a companion with a poorly constrained mass, are shown in Fig.~\ref{fig:gaiapmex_example}.

The results of our {\tt GaiaPMEX} analysis are provided in Table D.2. 
Given the Gaussianity of the single-star PMa$^{2/3}$ and UEVA$^{1/3}$ distributions, we adopted thresholds of $(\snr)_{\text{PMa}} = 3$ or $(\snr)_\ruwe = 3$ to identify possible binaries. We excluded from the statistical sample only the 39 stars for which the minimum mass of the companion was in the stellar regime (for instance, HIP 78851 in Fig.~\ref{fig:gaiapmex_example}). Stars where the {\tt GaiaPMEX} solution extended to substellar masses (for instance, HIP 3556 in Fig.~\ref{fig:gaiapmex_example}) were instead retained, not to induce a selection bias in our substellar demographics.

\subsection{Additional vetting criteria}
\label{sec:other_vetting_criteria}

As we are interested in deriving the properties of planets around young stars, we discarded  nine stars that after our reanalysis turned out to be older than 1\,Gyr. 
Furthermore, five stars with gas-rich circumstellar disks were removed, considering the possibility of ongoing planet formation and dust obscuration of any planet by circumstellar or circumplanetary disks, which may bias the statistical outcome of the survey, especially considering the very young age of these targets \citep[see, e.g.][]{cugno25}.

\section{Discussion}
\label{sec:discussion}

\begin{figure*}
    \centering
    \includegraphics[width=0.95\linewidth]{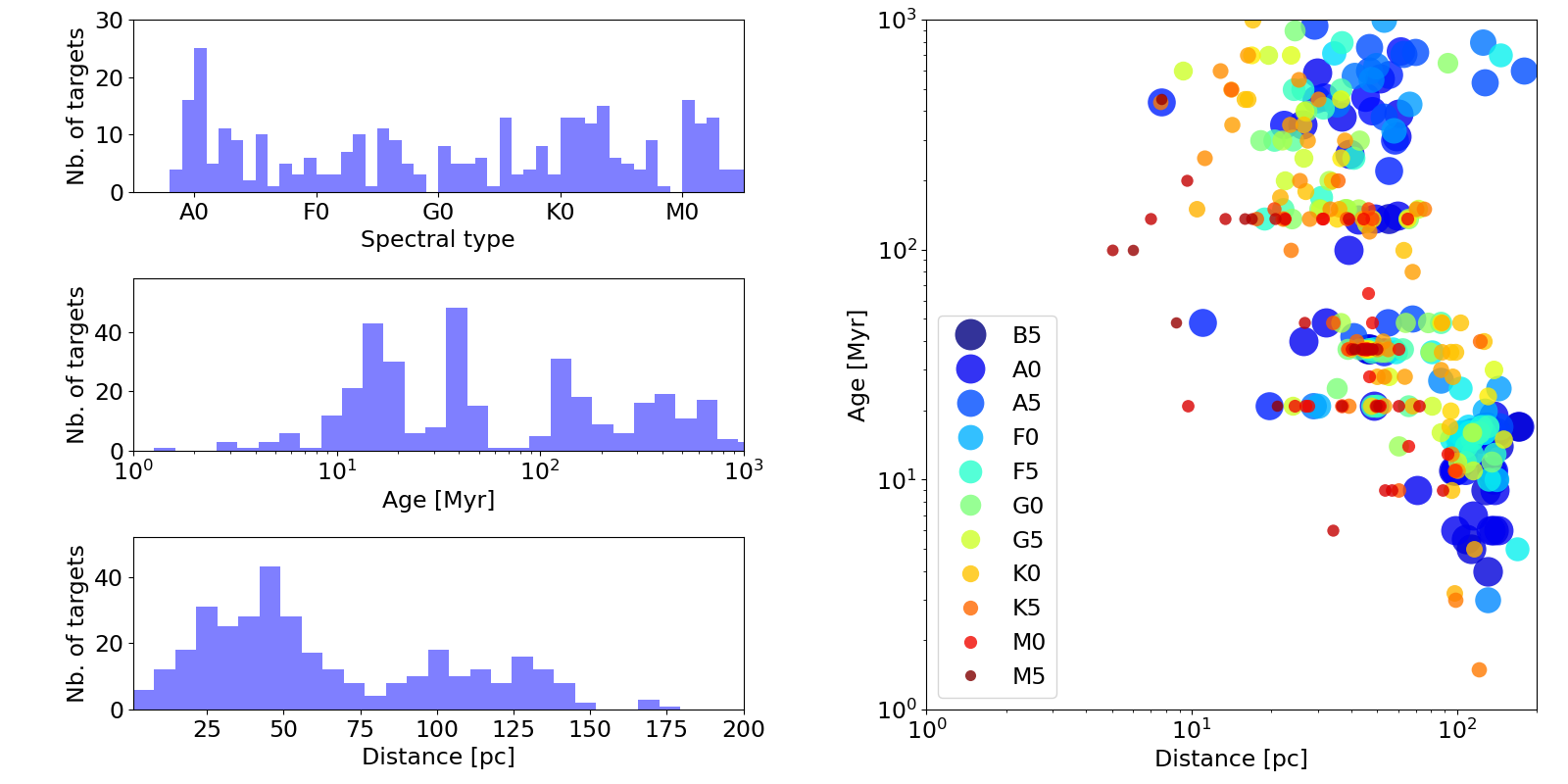}
    \caption{Left panel: Distribution of age, spectral type and distance for the 333 stars belonging to the statistical sample. Right panel: Relation among the three variables.}
    \label{fig:spt_age_distance}
\end{figure*}

The final statistical sample, reported in Table D.4, 
is composed of 333 stars. Fig.~\ref{fig:spt_age_distance} shows the distribution of age, spectral type and distance of this sample. The histograms highlight, in particular, the extensive coverage of the age and mass axes probed by SHINE. No dedicated study of metallicity was performed in this paper; instead, we assumed a solar metallicity based on the empirical finding that the metallicity of young star-forming region in the solar neighborhood is solar with a limited spread \citep{d'orazi11,biazzo12,baratella20,magrini23}.

\subsection{Global system architectures: Disks and additional planets}
\label{sec:disks_and_inner_planets}

Although a study of global system architectures lies outside the scopes of this work, it is nonetheless interesting to report the known planets and planet candidates around SHINE stars that are not detectable by SPHERE. To build this list, we cross-matched the F400 sample with the NASA Exoplanet Archive\footnote{\url{https://exoplanetarchive.ipac.caltech.edu/}.}, finding nine confirmed planets and  three planet candidates. The main properties of these planets, mostly discovered through transits or radial velocities, are listed in Table~\ref{t:inner_planets}. As the RV and transit methods preferentially target old stars, we emphasize that this census must not be considered complete.

\begin{table}[t]
\tiny
\caption{Confirmed and candidate inner planets in the SHINE sample.}
\centering
\begin{tabular}{llccccc}
\hline \hline
HIP                  & Pl.  &   $M_P$  & $R_P$ & $P$ & Technique &  Ref.  \\
 & & ($\Mjup$) & ($\Rjup$) & (d) &\\
\hline
\multicolumn{7}{c}{Confirmed planets} \\
\hline
102409  & b  &   $0.02 \pm 0.01$ &  $0.35 \pm 0.01$  & 8.5 &  T, TTV & 1\\
        & c  &   $0.04 \pm 0.01$ &  $0.23 \pm 0.02$  & 19 &  T, RV & 2 \\
        & d  &   $0.003 \pm 0.002$ & ...  &  13   &  TTV & 3 \\
71395  & b & $1.77 \pm 0.02$ & ...  & 453  &  RV & 5  \\
           & c & $3.8^{+0.9}_{-0.4}$ & ... & 922 &  RV & 6  \\
107412 & c & $13 \pm 1$ & $1.46^{+0.18}_{-0.06}$ & 2090 & A, INT, RV & 7 \\
27321 & c & $10 \pm 1$ & ... & 1190 & INT, RV & 8 \\
64184 & b & $8 \pm 1$ & $1.00 \pm 0.03$ & $110$ & T, RV & 9 \\
\hline
\multicolumn{7}{c}{Planet candidates} \\
\hline
102409 & e  &   $0.07(2)$ & ...  &  33   &  RV & 4 \\
64184 & c & ... & $1.15(5)$ & ... & T & 10 \\
56379 & b & ... & $9 \pm 4$ & ... & DI & 11 \\
\hline
\end{tabular}
\label{t:inner_planets}
\tablefoot{References: 1: \citet{plavchan20}; 2: \citet{martioli21}; 3: \citet{wittrock23}; 4: \citet{donati25}, 5: \citet{butler03}; 6: \citet{vogt05}; 7: \citet{hinkley23}; 8: \citet{lagrange19b}; 9: \citet{zakhozhay22}; 10: TESS; 11: \citet{quanz15}. Techniques: T: transits; TTV: transit-timing variations; RV: radial velocities; DI: direct imaging; INT: interferometry; A: astrometry.}
\end{table}

\noindent As mentioned in Section~\ref{sec:other_vetting_criteria}, five stars are surrounded by protoplanetary disks.  With respect to planetary disks, we cross-matched the F400 sample with the Catalog of Circumstellar Disks\footnote{\url{https://circumstellardisks.org/}.} and with the Spitzer Infrared Spectrograph Debris Disk Catalog \citep{chen14}; in addition to these archival sources, we also considered the recent analysis by \citet{engler25}, including, but limited to, the stars considered in this work. We count 46 resolved debris disks, 18 of which were resolved by SPHERE (the remaining ones are only detectable by instruments probing larger wavelengths such as ALMA). An additional 68 stars show evidence of a debris disk through infrared excess. These results are reported in Table D.1.

\subsection{Comparison with GPIES F400}

\begin{figure*}
    \centering
    \includegraphics[width=\linewidth]{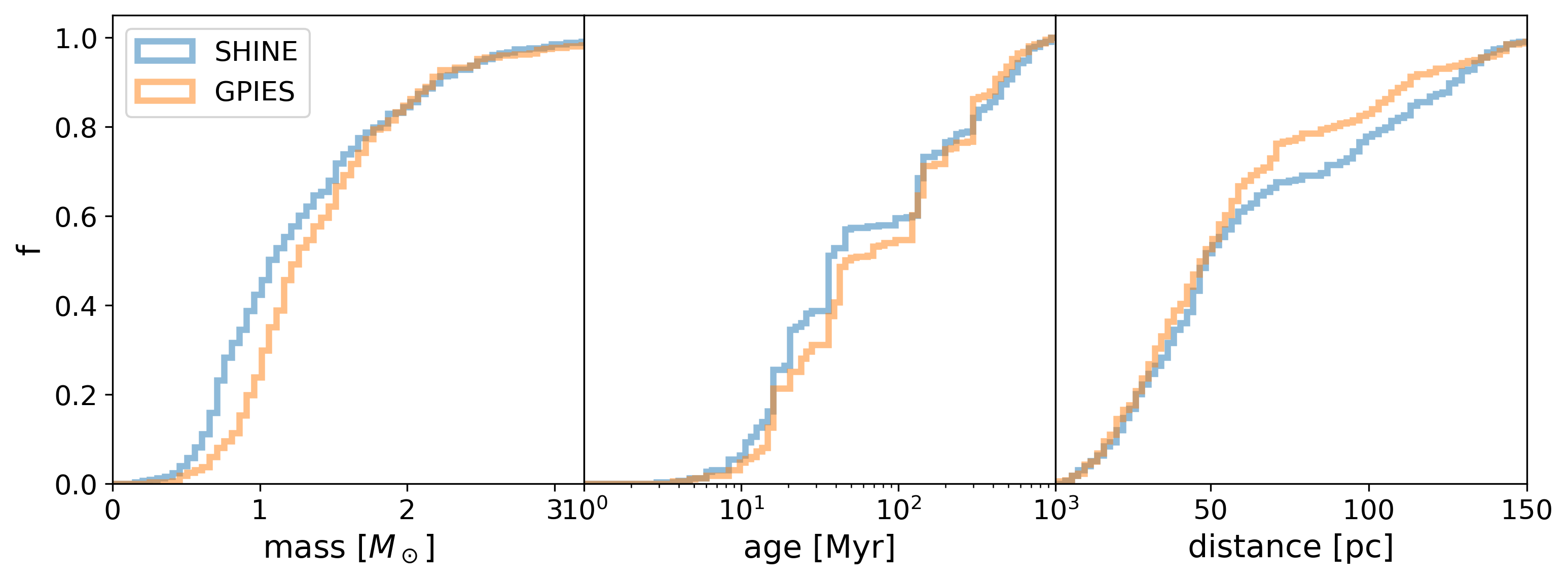}
    \caption{Cumulative distribution of mass, age, and distance for the stars in SHINE (blue, this work) and GPIES (orange, \citealt{squicciarini25}).}
    \label{fig:gpies_comparison}
\end{figure*}

We compared the properties of this sample with the one of GPIES \citep{nielsen19}. We start from the sample of 400 stars observed as part of GPIES and reanalyzed by \citet{squicciarini25}, a larger sample compared to the 300-star presented in \citet{nielsen19}. For the 180 stars common between SHINE and GPIES, we use the stellar parameters determined in this work. A comparison of the mass, age, and distance distribution of the two samples is shown in Fig.~\ref{fig:gpies_comparison}.

Compared to GPIES, SHINE benefits from a larger coverage of the low-mass stellar regime, thanks to SPHERE's capability to achieve a good AO correction for stars as faint as $R \sim 12$ mag \citep{beuzit19}. In GPIES (400 stars) we count 12 stars with $M_\star \leq 0.6~ \Msol$ and 26 stars with $ 0.6 < M_\star \leq 0.8~\Msol$, while in SHINE (333 stars) these numbers rise to 27 and 67, respectively.

SHINE stars tend, on average, to be younger and farther away than their GPIES counterparts, partially due to the higher fractional coverage of the Sco-Cen region (75/333 vs. 75/400) and the larger fraction of low-mass members of young moving groups. In this respect, we point out that the selection criteria of the GPIES F400 sample from the full population of GPIES targets are biased against intermediate-age field stars, implying that the age distribution of the full GPIES survey is even more dissimilar from the one of SHINE compared to what our plots suggest. Assuming that the typical detection limits achieved in the two surveys in the $H$ band are similar (cp. Fig. 2 from \citet{squicciarini25} and Fig. 4 from \citet{chomez25}), we might thus expect a somewhat deeper sensitivity of SHINE compared to GPIES in terms of minimum detectable planet mass. While this claim is currently difficult to assess, a second factor does certainly contribute to a deeper sensitivity across a wider range of semi-major axes: namely, the larger field of view of IRDIS ($11" \times 11"$) compared to GPI ($2.7" \times 2.7"$).

A certain difference exists in the treatment of binaries. Due to its aforementioned smaller field, GPIES includes multiples with a projected separation  $3" \lesssim \rho \lesssim 6"$ that are instead excluded in SHINE. In addition to this, no vetting for unresolved binaries was applied to GPIES \citep{nielsen19}. This affects the statistical results of that survey \citep[see, e.g., the recent discovery of the circumbinary HD 143811 b;][]{squicciarini25b,jones25}, making a direct comparison with SHINE challenging.

\subsection{Completeness to binary detection}

Despite the binary vetting procedure described in Section~\ref{sec:binaries}, some unresolved binary systems may still be present in the sample. In this section, we assess the level of residual contamination in the statistical sample starting from literature data on the occurrence and the properties of stellar companions.

We selected eight bona fide single stars (hereafter, {\tt GaiaPMEX} references) with no indications of binarity in the literature  or in {\tt GaiaPMEX}, chosen to sample the $(M_\star, d)$ parameter space defined by stellar masses of 0.5, 1.0, 1.5, and 1.9~$\Msol$ and distances of 50 and 100~pc. For each reference star, we injected stellar and substellar companions with masses in the range [1\,$\Mjup$, $M_\star$] and semi-major axes spanning [$10^{-2}$, $10^{3}$]~au, and computed the resulting astrometric solution (from which the $\snrruwe$ and $\snrpma$ defined in Section~\ref{sec:binaries} could be derived). This procedure allowed us to estimate the detection probability of astrometric companions as a function of companion mass and semi-major axis.

\begin{figure*}
    \centering
    \includegraphics[width=\linewidth]{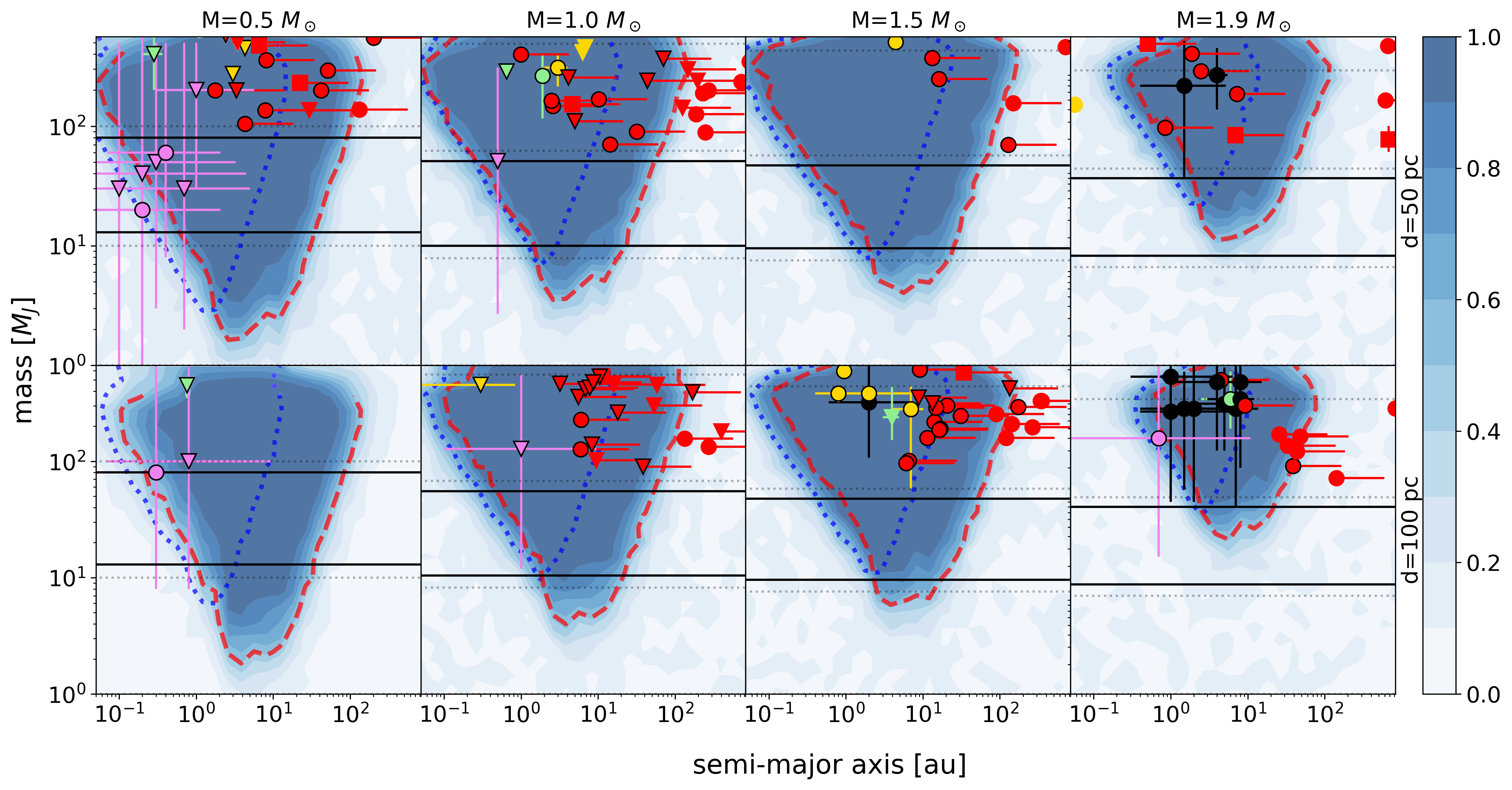}
    \caption{Completeness of our {\tt GaiaPMEX} search for binaries. Each panel represents one of the {\tt GaiaPMEX} references. The shaded area shows the completeness of the combined RUWE+PMa signal, with the dotted blue line and the dashed red line indicating the 50\% contour level given by RUWE and PMa alone, respectively. Overplotted are SHINE detections, each assigned to the closest reference star to its target. Squares indicate stars with no RUWE and PMa measurements; triangles represent stars without PMa; dots indicate stars with both RUWE and PMa. The companions are colored according to their detection method: red (DI), gold (RV), green (GaiaNSS), black ({\tt GaiaPMEX}); those also detected by {\tt GaiaPMEX} have a black edge. Finally, violet markers indicate companions, only detected by {\tt GaiaPMEX}, with a minimum mass in the substellar regime.}
    \label{fig:pmex_completeness}
\end{figure*}

The results of this analysis are shown in Fig.~\ref{fig:pmex_completeness}. To relate these simulations to the observed sample, each target star was associated with the closest {\tt GaiaPMEX} reference in the $(M_\star, d)$ plane. This enabled us to populate the figure with the companions identified in Section~\ref{sec:binaries}, which are shown using different colors and symbols depending on their detection method and on whether they were also identified by {\tt GaiaPMEX}. For directly imaged companions, projected separations were converted into semi-major axes using a statistical conversion factor appropriate for circular orbits, $1.15_{-0.15}^{+3.72}$ (95\% confidence level), following \citet{dupuy11}.

As expected, companions not detected by {\tt GaiaPMEX} predominantly reside at wider separations, where their astrometric signatures manifest primarily as linear trends. A rigorous assessment of the overall completeness would require detailed assumptions regarding the occurrence rate of stellar companions as a function of primary mass, as well as knowledge of the intrinsic distributions of companion masses and semi-major axes, which themselves are expected to depend on the mass of the primary star. Moreover, such an analysis would need to account for the fraction of stars with archival radial velocity measurements, which were used both in the survey design and during the vetting process to exclude spectroscopic binaries. In light of these complexities, we adopt a semi-quantitative approach aimed at determining whether the observed number of stellar companions is broadly consistent with expectations from the literature, or whether it is significantly lower, which would suggest a substantial residual contamination by unresolved binaries.

The mass-ratio distributions of stellar companions at different primary masses were taken from \citet{elbadry19}, adopting the values reported for separations in the range $50~\text{au} < s < 350~\text{au}$ owing to the lack of constraints at smaller separations. For the semi-major axis distribution, we followed the empirical trends presented in \citet{offner23}. We assumed, for simplicity, that the mass ratio and the semi-major axis are uncorrelated, which might not always be verified \citep[see, e.g.,][]{tokovinin14}.  For each {\tt GaiaPMEX} reference, we could therefore evaluate the product of the two distributions over a dense grid in companion mass and semi-major axis. The resulting distributions were normalized to reproduce the total stellar companion occurrence rates appropriate for each primary mass, again taken from \citet{offner23}. Integrating these distributions over a given region of the $(M, a)$ plane and multiplying by the number of stars associated with each {\tt GaiaPMEX} reference yields the expected number of stellar companions; applying the corresponding {\tt GaiaPMEX} completeness maps before integration provides an estimate of the number of companions detectable by {\tt GaiaPMEX}. We restricted this analysis to $a < 10$~au, since our SPHERE observations are expected to be largely complete for stellar-mass companions at wider separations, and considered the full stellar regime ($M > 80~\Mjup$).

The resulting mean {\tt GaiaPMEX} completeness, as anticipated, is higher for closer and lower-mass stars, ranging from approximately 60\% to 90\%. Based on the adopted empirical distributions, we expect a total of 72 stellar companions when summing over all {\tt GaiaPMEX} references, of which 59 should be detectable by {\tt GaiaPMEX}. In comparison, the F400 sample contains 72 binary stars, 61 of which are detected by {\tt GaiaPMEX}. An additional 13 companions are identified in our analysis but are not included here because their minimum masses fall in the substellar regime (Section~\ref{sec:binaries}).

Assuming Poisson statistics for the SHINE counts, we conclude that there is no evidence for a significant contamination of the statistical sample by unresolved stellar binaries. Given the numerous assumptions and uncertainties inherent in this analysis, any more quantitative conclusion cannot currently be drawn.

\section{Conclusions}
\label{sec:conclusion}
This paper presented an overview of the stellar sample targeted by SHINE, a large direct imaging survey looking for wide-orbit giant planets in the solar neighborhood. It belongs to a series of three papers comprising a study of the observations and data reduction \citep{chomez25}, and the statistical analysis emerging from the survey \citepalias{chauvin26}; these publications are an update of the papers \citep{desidera21,langlois21,vigan21} published when the survey was $\sim$30\% complete.

After providing updates on the observation setup and the survey execution with respect to the original design, we undertook a full reassessment of the stellar properties of the 460 stars targeted by the campaign. In particular, stellar ages were determined by means of a variety of methods: membership to young moving groups, activity, lithium, rotation, isochrone fitting. Stellar masses were derived by comparison with isochrones, given the corresponding ages.

We defined a subsample adequate for the statistical analysis to be performed in \citetalias{chauvin26}. The most important selection criterion was the exclusion of multiple systems with separations $s < 500$\,au. These were identified through the literature and a new astrometric analysis based on the {\tt GaiaPMEX} tool. We then quantified the completeness of {\tt GaiaPMEX} as a binary-vetting tool, and found with an empirical, semi-quantitative analysis that we do not expect a significant number of unresolved binaries to have eluded detection. The final statistical sample is made of 333 stars.

A comparison with the GPIES survey showed that the SHINE sample is better positioned to study the dependence of planet occurrence on stellar mass, owing to its larger number of faint K- and M-type stars, and is expected to probe more deeply the mass and semi-major axis space due to a slightly younger mean age and the large field of view of IRDIS. In the future, the combination of the two surveys will open up an even more powerful avenue for demographic efforts thanks to the unprecedented sample size.

\section*{Data availability}

The full version of the appendices, including Tables~D1--D5, are available on Zenodo: \url{https://zenodo.org/records/20312978}. Tables D1--D5 are also available at the CDS via anonymous ftp to cdsarc.cds.unistra.fr (130.79.128.5) or via \url{https://cdsarc.cds.unistra.fr/viz-bin/cat/J/A&A/}.

\begin{acknowledgements}

The authors would like to thank the anonymous referee for the insightful feedback that significantly helped improve the quality of the manuscript.

SPHERE is an instrument designed and built by a consortium consisting of IPAG (Grenoble, France), MPIA (Heidelberg, Germany), LAM (Marseille, France), LIRA (Paris, France), Laboratoire Lagrange (Nice, France), INAF - Osservatorio di Padova (Italy), Observatoire de Gen\`eve (Switzerland), ETH Z\"urich (Switzerland), NOVA (Netherlands), ONERA (France) and ASTRON (Netherlands) in collaboration with ESO. SPHERE was funded by ESO, with additional contributions from CNRS (France), MPIA (Germany), INAF (Italy), FINES (Switzerland) and NOVA (Netherlands). SPHERE also received funding from the European Commission Sixth and Seventh Framework Programmes as part of the Optical Infrared Coordination Network for Astronomy (OPTICON) under grant number RII3-Ct-2004-001566 for FP6 (2004--2008), grant number 226604 for FP7 (2009--2012) and grant number 312430 for FP7 (2013--2016). 

This work has made use of the High Contrast Data Centre, jointly operated by OSUG/IPAG (Grenoble), PYTHEAS/LAM/CeSAM (Marseille), OCA/Lagrange (Nice), Observatoire de Paris/LIRA (Paris), and Observatoire de Lyon/CRAL, and is supported by a grant from Labex OSUG@2020 (Investissements d’avenir - ANR10 LABX56).

This research has made use of the SIMBAD database and Vizier services, operated at CDS, Strasbourg, France. 

This research has made use of data obtained from The Extrasolar Planets Encyclopaedia.

This research has made use of the NASA Exoplanet Archive, which is operated by the California Institute of Technology, under contract with the National Aeronautics and Space Administration under the Exoplanet Exploration Program.

This research used the Washington Double Star Catalog, maintained at the U.S.\ Naval Observatory. 

This research has made use of the Exoplanet Follow-up Observation Program (ExoFOP; DOI: 10.26134/ExoFOP5) website, which is operated by the California Institute of Technology, under contract with the National Aeronautics and Space Administration under the Exoplanet Exploration Program.

This work has made use of data from the European Space Agency (ESA) mission
{\it Gaia} (\url{https://www.cosmos.esa.int/gaia}), processed by the {\it Gaia}
Data Processing and Analysis Consortium (DPAC,
\url{https://www.cosmos.esa.int/web/gaia/dpac/consortium}). Funding for the DPAC
has been provided by national institutions, in particular the institutions
participating in the {\it Gaia} Multilateral Agreement.

Based on data retrieved from the SOPHIE archive at Observatoire de Haute-Provence (OHP), available at \url{http://atlas.obs-hp.fr/sophie/.}

V.S.\ acknowledges support from the UK Science and Technology Facilities Council (STFC) under Grant Code ST/Y002792/1.

G.-D.M.\ acknowledges the support of the Deutsche Forschungsgemeinschaft (DFG) through grant MA~9185/2-1.

\end{acknowledgements}

\bibliography{f400}
\bibliographystyle{aa_url}

\begin{appendix}
\section{Notes on individual objects in the statistical sample: F150 targets}

\label{a:notes_f150}

\begin{description}

\item {\it HIP 54231 = HD 96338.}
As in \citetalias{desidera21}, we employed an upper age of 300\,Myr, derived from isochrone fitting, to account for the uncertain membership to LCC.

\item {\it HIP 77457 = HD 141190.}
As in \citetalias{desidera21} and similarly to the case above, we combined the YMG age and isochronal results to account for a disputed membership to USCO. However, we adopted here the isochrone result (800\,Myr) as nominal age.

\item {\it HIP 78541 = HD 143488.}
As for HIP 54231, we employed an upper age of 300\,Myr to encompass the possibility of a  nonmembership to UCL.

\item {\it HIP 107412 = HD 206893.}
In addition to the known HD 206893 B \citep{milli17b}, a second companion was identified by \citet{hinkley23}. The two companions are responsible for the significant PMa that is observed.

\item {\it HIP 113283 = HD 216803.}  
Second component of the Fomalhaut (HIP 113368) system. 
Age from \citet{mamajek12}.

\item {\it TWA 6 = GSC7183-1477 = BX Ant.}
\banyan~ with Gaia DR3 parameters does not support membership to TWA, in agreement with the DR2-based analysis discussed in Paper I.
The star is also not considered a TWA member in recent studies \citep{luhman23,miretroig25}.
There are some indications of binarity: RUWE=1.415, corresponding to $\snrruwe=3.4$ according to {\tt GaiaPMEX}; scatter in RV measurements in the literature.
The very large EW(Li) \citep{sacy} is in any case a robust indication of an age younger than BPIC. Based on isochrone fitting, we adopted $t_\star=14 \pm 1$\,Myr.

\item {\it TYC 9482-121-1.}  
Field star of intermediate age \citep{desidera21} with a significant RUWE ($\snrruwe=7.3$). The star has AstroSpectroSB1 solution in Gaia-NSS, with nominal mass in the substellar regime \citep{marcussen2023}. This is confirmed by our independent {\tt GaiaPMEX} analysis; therefore, we retained it in the statistical sample.

\end{description}

\section{Notes on individual objects in the statistical sample: new targets}

\label{a:notes_f400}

\begin{description}

\item {\it 2MASS J10252092-4241539.}   
M1 star proposed as TWA member by \citet{bell15} and \citet{luhman23}.
The kinematic analysis with Gaia DR3 astrometric parameters and RV from \citet{malo14}
yields null membership probability.
The recent analysis of TWA by \citet{miretroig25} assigned the star to the subgroup TWA-a. 
Independently of the kinematic assignment, the star is very young: the EW(Li) \citep{malo13} indicates an age younger than the one of the BPIC and compatible with TWA.
With respect to multiplicity, the star is classified as EB by \citep{parihar09} and as a rotational variable by \citet{kiraga12}. However, it is not classified as a binary in TESS papers. The RUWE of this star is 1.34 ($\snrruwe=2.8$), leaving the question open. We tried to fit the properties of the system assuming that the EB classification is correct: based on the reported transit depth, we estimated $\Delta V \sim \Delta G=0.66$ mag. This yielded an age of $\sim 33$\,Myr. However, such an age is not compatible with lithium. Conversely, the single-star assumption yielded an age of $\sim 10$\,Myr consistent with this indicator. We interpret the photometric modulations as due to rotational variability, rather than multiplicity.

\item {\it 2MASS J16430128-1754274.}   
Active  M0.5 dwarf, classified as a member of the BPIC in several papers in the literature starting from \citet{kiss11}.
However, the Gaia DR3 kinematic parameters yield a null membership probability using \banyan. 
Membership was also rejected in a recent study by \citet{lee24}.
A young age, fully compatible with BPIC, is supported by lithium and rotation period
\citep{messina16,Messina17}.
The position on the color-magnitude diagram (CMD) well above the zero-age MS (ZAMS) is indicative of a very young age. 
It is possible that the kinematic parameters and the CMD position are somewhat altered by binarity. The star has a RUWE on the verge (RUWE=1.32, $\snrruwe)=2.9$) and a large scatter in {\it Gaia} RV, possibly indicating the presence of a companion at separation closer than the sensitivity of SPHERE observations.
However, the three higher quality RV measurements by \citet{malo14} show a small scatter ($=0.5$ km/s) over a baseline of four years and rule out an SB2 system.
We consider the star only as a suspected binary, and adopted an age $t_\star=23 \pm 3$\,Myr from isochrone fitting. The uncertainties on stellar parameters encompass a $q$ up to 0.2.

\item {\it 2MASS J23221088-0301417.}   
Active and fast rotating K7 star, flagged as BPIC or Columba candidate by \citet{naud17}. However, \banyan~yields a null membership probability and \citet{bowler19} did not detect lithium, expected to be prominent at such young age.
The position on CMD is also close to ZAMS, but distant enough to allow a tight age constraint from isochrones ($t_\star=65 \pm 5$\,Myr).
There are no robust evidences of a tidally-locked companion. 
We adopted the age determination from isochrone fitting, which would be compatible with the observational constraints listed above.

\item {\it AD Leo = GJ 388.}  
Nearby (d $\sim$ 5 pc) and active M dwarf (SpT = M3.5V), widely studied as benchmark for various magnetic activity phenomena \citep[see, e.g.,][]{hawley03}.
The controversial close-in planet was dismissed in the most recent literature as due to activity \citep{carleo20,kossakowski22,carmona23}.
Given the peculiarity of this star, in addition to the standard PARSEC model (yielding $t_\star=100_{-5}^{+10}$\,Myr), we also tested the magnetic Dartmouth grids \citep{feiden16}. The latter indicate a slightly younger age between 80 and 100\,Myr. Hence, we adopted a looser $t_\star=100_{-20}^{+10}$\,Myr.

\item {\it GSC 7396-0759.}  
M1V star, it is a wide companion (170") to the spectroscopic binary V4046 Sgr. Spatially resolved debris disk \citep{sissa18,cronin22}. Bona fide member of BPIC according to our \banyan~analysis.

\item {\it HIP 9716 = HD 12786 = FN Cet.}   
Field K0V star, classified as possible member of the Castor moving group \citep{vereshchagin15}.
The rotation period \citep{fetherholf23}, the X-ray emission, the chromospheric emission \citep{wright04} indicate an age intermediate between Hyades and Pleiades, similar to the proposed one for Castor regardless of the real existence of the group.
The longer rotation period (about twice the TESS one) by \citet{strassmeier00} is rejected, as it is not compatible with the observed \vsini~and the expected one from $\log{R'_{\text{HK}}}$.
Our measurement of EW(Li) from SOPHIE spectra is consistent with the age from the other methods.
We adopted $t_\star=350 \pm 100$\,Myr.

\item {\it HIP 17439 = HD 23484 = GJ 152.}   
Nearby ($d=16$ pc) K2V star with a spatially resolved debris disk \citep{ertel14}.
The star is flagged as SB in \citet{gaidos98} but there are no evidences of RV variability from \citet{nordstrom04}, CORALIE \citep{soubiran18} and {\it Gaia}. Furthermore, there is no significant excess astrometric scatter in {\it Gaia}.
$\log{R'_{\text{HK}}}$ and the (uncertain) rotation period are similar to those of the Hyades, and the X-ray emission is a bit lower than the Hyades locus. The non detection  of lithium is consistent with this age assignment.
We adopted $t_\star=700_{-150}^{+200}$\,Myr.

\item {\it HIP 28764 = HD 41700.}
F8V star, part of a quintuple system.
The star is flagged as SB in SIMBAD but this appears to be due to an exchange of components.
Indeed, \citet{tokovinin16} showed that the SBs in the system
are the two components of the pair ($\rho = 5"$) HIP 28790 (HD 41742A
and HD 41742B). Our target HIP 28764 = HD 41700 is a comoving object at 196.1" = 5300\,au from HD 41742A and has roughly constant RV and no significant astrometric scatter.
From the age indicators of HD 41700 ($P_{\text{rot}}$, $\log{R'_{\text{HK}}}$,  EW(Li)) and the nondetection of lithium in the spectra of K4.5 component HD 41742B \citep{tokovinin16}, we inferred an age of $400 \pm 200$\,Myr.

\item {\it HIP 28910.}
A0V star, which \banyan~ labels as a field star -- albeit with a low (4.4\%) membership probability to Argus. Our isochronal analysis argued against this residual possibility, yielding $t_\star=555 \pm 40$\,Myr.

\item {\it HIP 28921 = HD 41842.}
K1V star with a wide companion (19" = 690\,au), whose physical association to our target is confirmed by {\it Gaia}. The age of the primary, a field star according to \banyan, is virtually unconstrained by isochrones. 
The companion, 2MASS J06061742-2754050, is a M3Ve star with RUWE=1.3, $\snrruwe=2.5$. If this CMS is not in turn a binary, then its CMD position argues for $t_\star \sim 100$\,Myr. In the extreme case of an equal-mass binary, it would instead be consistent with $t_\star = 300-400$\,Myr. The single-star scenario for 2MASS J06061742-2754050 provides an age that is comparable to the one estimated by \citet{nielsen19} (150\,Myr). We conservatively assumed $t_\star=150_{-50}^{+250}$\,Myr.

\item {\it HIP 28954 = HD 41593 = GJ 227 = V1386 Ori.}
Nearby ($d=15$ pc) K0V star with ambiguous membership to UMa from several sources \citep{montes01,king03,dopcke19}; however, the membership probability from \banyan~ is null.
All age indicators ($P_{\text{rot}}$, \citealt{capistrant24}; $\log{R'_{\text{HK}}}$, \citealt{wright04}; $R_X$, \citealt{ammler09}) are in any case fully compatible with UMa age. We therefore assumed the YMG age, $t_\star=450 \pm 100$\,Myr.

\item {\it HIP 32938.}
A3V star classified as a field star by \banyan. 
It has an M-type wide companion at 20" = 1100\,au, Gaia DR3 5578901662668399872.
Isochronal analysis for the primary suggested either $t_\star \sim 10$\,Myr or $t_\star \sim 600$\,Myr. The position of the CMS on the CMD was crucial to rule out the young solution. We adopted $t_\star = 580_{-40}^{+45}$\,Myr.

\item {\it HIP 34782.}
A8III/IV star classified as a field star by \banyan. 
It has an M-type wide companion at 25" = 1200\,au, Gaia DR3 5605930265526073472, showing clear hints of binarity (RUWE $=$ 5.1, $\snrruwe=40$). Isochronal analysis for the primary suggests either $t_\star \sim 15$\,Myr or $t_\star \sim 600$\,Myr.
Despite the degeneracy in the ($q$, $t_\star$) space, the CMS allowed us to rule out the young solution -- its photometry is not compatible with $t_\star < 50$\,Myr even in the limit case of a single star. Values $q \approx 1$ are needed to reconcile the ages of the two components. We adopted $t_\star=550_{-60}^{+65}$\,Myr.

\item {\it HIP 35884 = HD 58192.}
F7V star, whose adopted age is based on indirect indicators.
The X-ray emission is similar to the value of Hyades members, $\log R'_{HK}$ intermediate between Hyades and Pleiades, the photometric period \citep{fetherholf23,colman24} is at the edge
of available calibrations and, considering empirical loci, above Group X
\citep{messina22} and below Praesepe.
We then adopted 500$\pm$200\,Myr. 

\item {\it HIP 36515 = HD 59967.} 
The age from isochrone fitting ($t_\star=300_{-240}^{+300}$\,Myr) is broadly consistent
with the outcome of indirect indicators \citep[\prot $= 5.18$~d;][]{fetherholf23}
\citep[$\log R'_{HK} = -4.42$ dex;][]{lorenzooliveira18} \citep[A(Li) $=$ 2.81;][]{ramirez12}, which suggest a lower limit around 150\,Myr. We adopted $t_\star=300_{-150}^{+300}$\,Myr.

\item {\it HIP 36827.}
Field K2 dwarf with IR excess \citep{cotten16}.
Age $350_{-100}^{+150}$\,Myr from rotation, lithium, X-ray and chromospheric emission.

\item {\it HIP 37288.}
M0V star, field star according to \banyan. Its level of chromospheric activity \citep[$\log{R'_{\text{HK}}}=-4.53$ dex;][]{shan24} and its rotational period from \vsini~ \citep[$P=15.30$ d;][]{reiners18} suggest an age $t_\star=1.0 \pm 0.3$\,Gyr.

\item {\it HIP 37349 = HD 61606A = V869 Mon.}
Very nearby ($d=14$ pc) wide binary system with HD 61606B ($\rho = 57.9" \sim 815$\,au). Both components were included in the sample and observed in the SHINE survey. Age from \citet{vigan17}.

\item {\it HIP 40706.}
Field early type (A8V) star.
We adopted $t_\star=945 \pm 35$\,Myr from isochrone fitting.

\item {\it HIP 41282.}
F2V star, with X-ray emission and activity, which however have limited sensitivity to age at the target \teff.
The kinematic parameters are slightly outside the kinematic box by \citet{montes01} but still compatible with a moderately young star. Based on isochrone fitting, we estimated $t_\star=715_{-170}^{+220}$\,Myr.

\item {\it HIP 41889 = FR Cnc.}
Fast-rotating K7 star \citep{golovin12}.
Detection of lithium, below the mean value of Pleiades members of similar color, but within the scatter, indicate a young age.
Kinematic properties are somewhat ambiguos, with 79.7\% probability of a field object, but a nonnegligible 16.8\% and 3.5\% probability for memberhip in Argus and Carina-Near, respectively. Isochrone fitting favors either $75 ~\text{Myr}<t<90~\text{Myr}$, or $130 ~\text{Myr}<t<500~\text{Myr}$. We conservatively adopt $t_\star=200_{-125}^{+300}$\,Myr.

\item {\it HIP 41967 = HD 72687.}
Its kinematics are similar to the group of stars comoving with TOI-1807 and TOI-2046, but significantly younger constraints were obtained from lithium and rotation period \citep{nardiello22}. We adopted $t_\star=130_{-50}^{+70}$\,Myr.

\item {\it HIP 42333 = HD 73550 = V401 Hya.}
Well-known active G5V star with discrepant measurements in the literature.
We checked the EW(Li) measurement, which results 35.0$\pm$1.5 m\AA from our analysis of HARPS archive spectra.
This is below the value of Hyades stars of similar color.
There are various measurements of rotation period ranging from 5.85~d \citep{tu22} to 12.4~d \citep{vidotto14}, with intermediate values also present \citep[$P=8.07$~d, ][]{fetherholf23}.
X-ray emission is close but slightly below the Hyades value, supporting the EW(Li) age and the longer rotation period.
The latter is compatible with \vsini. The shorter ones are also compatible, provided that the star is seen at intermediate inclinations $\sim$ 30 deg.
We adopted $t_\star=900$\,Myr with a lower limit at 300\,Myr to accommodate to the possibility of fast rotation and discrepant lithium from unknown reasons. 

\item {\it HIP 43290 = HD 75519.}
G8V star, for which the indirect methods (rotation period, $\log R'_{HK}$, X-ray emission, EW(Li)) yielded an age of $t_\star=250_{-70}^{+100}$\,Myr. The results of isochrone fitting are fully consistent with this age range.

\item {\it HIP 43797 = HD 76653 = GJ 3519.}
F6V star with IR excess \citep{chen14} . The combination of indirect methods (EW(Li), X-ray emission, $\log{R'_{\text{HK}}}$, kinematics) yielded a most likely age around 500\,Myr, with lower and upper limit at 300\,Myr and 2\,Gyr. The isochronal analysis indicated a tighter $t_\star=800 \pm 300$\,Myr, that we adopted as our age estimate.

\item {\it HIP 46580 = HD 82106 = GJ 349.}
Nearby ($d=12.8$ pc) K3 dwarf.
There is some ambiguity on the rotation period. The determination by \citet{fetherholf23} from TESS data ($P=5.846$ d) would imply a gyrochronology age of about 160\,Myr, in tension with the activity level, the non detection of lithium and \vsini~ (it would imply an orientation close to pole-on). A period twice the proposed one would nicely fit the results of the other indicators. An even longer rotation period (13.3 d) was derived by \cite{saar1997}. We adopted, following \citet{vigan17}, $t_\star=600 \pm 200$\,Myr.

\item {\it HIP 46709.}
Field early type (A9IV/V) star. 
We adopted $t_\star=330_{-70}^{+80}$\,Myr from isochrone fitting.

\item {\it HIP 49127 = HD 86972.}
K3 dwarf. 
Kinematic analysis yields a 30.5\% membership probability to CARN.
The rotation period by \citet{fetherholf23} indicates an age of 160\,Myr.
Such an young age could be consistent with the possible Carina-Near membership, but it is in tension with our marginal lithium detection (EW(Li) $ = 2.8\pm 1.0$ m\AA),
the expected rotation period from $\log R'_{HK}$ \citep{wright04}, and the X-ray nondetection. A true rotation period two times the observed one with TESS would
be in better agreement with the various indicators. We then adopted an age of 550\,Myr with a lower limit at  150\,Myr to include the nominal rotation period and possible Car-Near membership and upper limit at 700\,Myr.

\item {\it HIP 50083.}
A4mA6-F0 star classified as Sco-Cen member (LCC) by \citet{dezeeuw99}, but the membership is rejected by our kinematic analysis. The isochronal analysis yielded two degenerate solutions: ($t_\star \sim 2$\,Myr, $M_\star \sim 2.9~M_\odot$) or ($t_\star = 600\pm100$\,Myr, $M_\star = 2.45 \pm 0.05~M_\odot$). Lacking any indications of youth in the literature, we opted for the latter solution, which is probabilistically more likely.

\item {\it HIP 51271 = HD 90884 = TOI-1028.}
K3.5V star, around which TESS identified a transiting planet candidate (TOI-1028.01). However, follow-up observations showed that the eclipses are occurring on an unrelated object at about 40" (Gaia DR3 5357886837517697408) \citep{guerrero21}, which was also identified as an EB in GaiaNSS (with 2$\times$ the TESS period).
The isochronal analysis indicates an age of 100\,Myr, with a lower limit at 80\,Myr and a tail extended up to 150\,Myr.
The outcome of indirect indicators is somewhat ambiguous.
The rotation period by \cite{fetherholf23} is in nice agreement with the isochrone age. 
However, the photometric period by \cite{canto20}, which is roughly two times
the one by \cite{fetherholf23}, $\log{R'_{\text{HK}}}$, and X-ray emission would
favor an older age (about 400\,Myr). 
The EW(Li) from CHIRON spectra acquired as part of the TESS follow-up\footnote{The spectra are available on EXOFOP at \url{https://exofop.ipac.caltech.edu/tess/target.php?id=447283466}, PI S. Quinn.} results 69.6$\pm$1.3 m\AA. This indicates a nominal age of 150-200\,Myr but is within the available measurements of Pleiades members and not compatible with the expectations for a 400\,Myr star.
Considering the lack of evidence for binarity and the constraints from Li, we adopted an age $t_\star=120 \pm 40$\,Myr.

\item {\it HIP 51386 = HD 90905.}
We adopted $t_\star=170_{-70}^{+180}$\,Myr following \citet{vigan17}.

\item {\it HIP 52462 = HD 92945   =    V419 Hya.}
Star with debris disk \citep[see][]{golimowski11,lazzoni25}.
Adopted age from combination of indirect indicators. See \cite{mesa21} for details. 

\item {\it HIP 53771.}
A3III/IV star with a wide companion, Gaia DR3 5359955053246566144 at 10" = 580\,au. Classified as a member of LCC by \citet{baron19}, it is a field object according to the \banyan~ analysis (although with a 30\% membership probability to Carina).
The M-type companion appears to be $\sim 50$\,Myr according to all the tested models but Dartmouth magnetic, according to which it can be compatible with any age up to 1\,Gyr. The isochronal analysis for the primary argues for an age around 250-300\,Myr, with an upper age limit at about 500\,Myr; the age can decrease to $\sim 50$\,Myr if $E(B-V)=0.05$  mag. We conservatively adopted a nominal age of 300\,Myr with lower and upper limits at 50\,Myr and 500\,Myr respectively.

\item {\it HIP 53824.}
Field early type (A5III) star.
We adopt $t_\star=760 \pm 50$\,Myr from isochrone fitting.

\item {\it HIP 54688.}
Field early type (A5V) star.
We adopted $t_\star=710_{-50}^{+60}$\,Myr from isochrone fitting.

\item {\it HIP 56227.}
F0III/F0IV star classified as a Sco-Cen member in \citet{pecaut12} but rejected by \banyan~ ($p=1\%$). The analysis by \citet{chen11} suggested an age compatible to Sco-Cen based on a variety of indicators (IR excess, EW(Li), \vsini, $\log{R'_{\text{HK}}}$). Both the photometric variability period from TESS, $P_{\text{TESS}}=0.41$~d \citep{fetherholf23}, and the RV variability of about 1 km/s \citep{grandjean23}, are not conclusive as they might be due to pulsations. We adopted a young age compatible with Sco-Cen, with an upper age of 600\,Myr from isochrones to account for a possible nonmembership.

\item {\it HIP 56543.}
A5V star, listed as Sco-Cen member in \citet{dezeeuw99} but rejected by \banyan~ ($p < 1\%$). Lacking any measurement of youth indicators in the literature, we assumed the MS isochronal age as baseline value, employing a minimum age of 7\,Myr (derived from isochrones as well) to account for a possible Sco-Cen membership.

\item {\it HIP 60561.} 
Our \banyan~ analysis argues for membership to EPSC rather than Sco-Cen as reported by \citet{luhman22}. The usage of the subsolar metallicity [Fe/H]$=-0.2$ -- coming from the Gaia-Apsis analysis \citep{gaia22} -- was required in the isochrone fitting to find an agreement between the EPSC age and the observed photometry.

\item {\it HIP 60965.}
A0IV star, classified as a field star by \banyan. Only loose constraints can be obtained from isochrone fitting, yielding a best-fit age of 40\,Myr with lower and upper ages equal to 9\,Myr and 200\,Myr. An early K-type CMS, Gaia DR3 3520585968137789184, exists at 23" $= 600$\,au, but did not provide additional age constraints.

\item {\it HIP 62703.}
A5V star traditionally assigned to LCC \citep[see, e.g.,][]{damiani19}. Its RV $= -6.9 \pm 2.0$ km/s is incompatible with this membership (according to \banyan, we would expect RV $= 12.4 \pm 4.3$ km/s); however, the measurement ultimately comes from \citet{andersen83} and must therefore be taken with caution. In any case, the CMD position of the star is too distant from the 20\,Myr isochrone -- even if considering the extreme case of an equal-mass binary. We adopted $t_\star=535 \pm 15$\,Myr from isochrone fitting.

\item {\it HIP 65178.}
B9V star, bona fide member of LCC according to our \banyan~ analysis.
A larger reddening than the one suggested from our maps -- $E(B-V) \approx 0.08$ mag, in agreement with recent studies \citep{gontcharov17,paunzen24} -- is needed to reconcile the photometry of the star and the group age.
An alternative scenario we investigated is that of a binary system with $q>0.6$ ($M \gtrsim 1.6 M_\odot$). However, no indication of such a prominent companion was found in the literature; nondetection by DI \citep[][this work]{kouwenhoven05}, TESS, and {\it Gaia} (RUWE and PMa) ruled out most of the parameter space, and the agreement between the two existing RV measurements \citep{gontcharov06,gaiadr3} made the possibility of a tight binary system even less likely. 
We opted thus for the first alternative.

\item {\it HIP 66068.}
A1/2V star, bona fide member of UCL according to our \banyan~ analysis.
A larger reddening than the one suggested from our maps -- $E(B-V) \approx 0.08$ mag, in agreement with \citet{gontcharov17} -- is needed to reconcile the photometry of the star and the group age.
The alternative hypothesis of unresolved binarity, for which there is no hint from {\tt GaiaPMEX} or from the literature, leads to a very high $\chi^2$ when coupled with the YMG age and was therefore rejected.

\item {\it HIP 67973.}
B9V star, bona fide member of LCC according to our \banyan~ analysis.
A larger reddening than the one suggested from our maps -- $E(B-V) \sim 0.11$ mag, in agreement with \citet{gontcharov17} -- is needed to reconcile the photometry of the star and the group age.
The alternative hypothesis of unresolved binarity, for which there is no hint from {\tt GaiaPMEX} or from the literature, can be led to agree with the YMG age only on condition of a very high $\chi^2$ and was therefore excluded.

\item {\it HIP 71395 = HD 128311.}
K3V star with two confirmed giant planets \citep{butler03,vogt05} and a planet candidate from RV and a long term RV trend. No significant updates on age indicators after \citet{vigan17}.
The object at 5.8" listed in WDS is a background star.

\item {\it HIP 73266.}
B9V star, bona fide member of UCL according to our \banyan~ analysis.
A larger reddening than the one suggested from our maps -- $E(B-V) \sim 0.07$ mag, in agreement with \citet{gontcharov17} -- is needed to reconcile the photometry of the star and the group age.
The alternative hypothesis of unresolved binarity with $q \sim 0.3$ does not find any correspondence in {\tt GaiaPMEX} or in the literature.

\item {\it HIP 74865.}
The observed PMa is due to the known BD companion \citep{Hinkley15}.
2MASS J15174874-3028484 at 95" is probable very wide companion
\citep{majidi2020}.

\item {\it HIP 76395.}
The star is classified as a close, equal-mass binary ($\rho = 0.2"$, $\Delta V \sim 0$ mag) in the Washington Double Star Catalog \citep{mason01}. The reported companion was observed twice in 1926 and 1933, but it has never been redetected since. We found no hint of stellar companions in SPHERE science and PSF images, ruling out any equal-mass companion up to $\sim 40$ mas. Astrometric constraints from {\tt GaiaPMEX} are even tighter, excluding such a companion up to $\sim 10$ mas. Therefore, we considered the archival detection as spurious.

\item {\it HIP 79258.}
F3/F4V star, indicated as an US member in several works
\citep{dezeeuw99,pecaut12,galli18, damiani19} but rejected as such by \banyan~ ($p<1\%$). Youth indicators reported by \citet{chen11} do not argue for a very young age. Slow rotator from HARPS spectra. A wide M-type companion (Gaia DR3 6035727063034449408) at $22"=3200$\,au further reinforces this scenario, as it is not overluminous with respect to the ZAMS. Based on isochrone fitting, we adopted $t_\star=700_{-500}^{+300}$\,Myr.

\item {\it HIP 79881.}
A1V star with controversial membership to BPIC in the literature \citep[see][]{zuckerman04,torres08,malo13,alonsofloriano15,bell15}.
\banyan~ with Gaia DR3 parameters yields a null membership probability. It should be noted that the errors are fairly large as the star is very bright ($V=4.78$  mag). This also explains why the high RUWE$=$1.62 is not significant ($\snrruwe=1$). IR excess noted by \citet{cotten16}.
Based on isochrone fitting, we adopted $t_\star=135_{-45}^{+65}$\,Myr.

\item {\it HIP 81935.}
 K3V star with rotation period equal to 8.536 d \citep{fetherholf23}, similar to Group X.
The nondetection of lithium \citep{sacy} is confirmed by our own analysis of an archive FEROS spectrum and ruled out ages younger than 200\,Myr.

\item {\it HIP 85038 = HD 156751.}
Visual binary ($\rho = 9.3" \sim 630$\,au). The primary, an A5V star, has sparse RV monitoring with HARPS. The astrometric signal from PMa is not significant ($\snrpma=2.6$) according to {\tt GaiaPMEX}; the combination of the PMa signal, the low RUWE and the constraints from DI rule out any stellar companion at all separations.
The secondary has colors corresponding to a late G or early K star 
and is an astrometric binary from the large RUWE (12.54) and likely just below the spatial resolution of {\it Gaia}
(ipd\_frac\_multi\_peak = 47) making the system triple.
The X-ray source 1RXS J172244.6-582829 is likely associated with HD 156751B, considering the spectral type of the primary.
Assuming this is the case, the X-ray emission result similar to Pleiades stars and larger than the Hyades, although with larger uncertainties because of multiplicity.
This indirect indicator would corroborate the age estimate obtained through isochrones, $t_\star=50_{-30}^{+150}$\,Myr. The unresolved multiplicity of HD 156751B makes it impossible to further constrain the age of the system.

\item {\it HIP 85922.}
A5V star classified as a field star by \banyan. Lacking any indirect age indicator, we estimated $t_\star=600\pm50$\,Myr based on isochronal analysis.

\item {\it HIP 86672.}
Young, active  G5V star with possible membership to BPIC (58.8\% with Gaia DR3 parameters). The membership is rejected by \citet{crundall19}.
A very young age is clearly supported by the strong lithium and activity.
Also, the CMD position is well above ZAMS, indicating a pre-MS evolutionary phase. Our new isochronal analysis led to a younger age ($t_\star=16 \pm 1$\,Myr) than the one reported by \citet{vigan17}. See \citet{desidera15} for further details on the object.

\item {\it HIP 87108.}
A1V star, classified as a field star by \banyan. Lacking any indirect age indicator, we estimated $t_\star=590 \pm 50$\,Myr based on isochronal analysis.

\item {\it HIP 87174.}
F0IV star classified as a field star by \banyan.
The isochronal analysis returned two solutions: $t_\star = 10 \pm 1$\,Myr or $t_\star = 1000 \pm 150$\,Myr. Lacking any indications of youth in the literature, we opted for the latter solution, which is probabilistically more likely.

\item {\it HIP 87836.}
A7III/IV star classified as a field star by \banyan. Lacking any indirect age indicator, we estimated $630 \pm 50$\,Myr based on isochronal analysis.

\item {\it HIP 89728 = HD 168159.}
K3V star, for which we adopted an age of 450$\pm$200\,Myr from the analysis of the available indirect indicators
(our measurement of  EW(Li), 24.9$\pm$8.0 m\AA; $\log{R'_{\text{HK}}}$ from \cite{gomesdasilva21}
and ROSAT X-ray emission. The photometric rotation period is not available.

\item {\it HIP 90133.}
A0V star, field star according to \banyan. Lacking any indirect age indicator, we estimated $t_\star=310\pm40$\,Myr based on isochronal analysis.

\item {\it HIP 90899.}
Star with a somewhat uncertain G0/2V classification \citep{skiff09}.
\banyan~ labels it as a field star -- albeit with a moderate (32.5\%) membership probability to THA. No indirect age indicator from the literature. Our isochronal analysis did not allow for a tight age constraint ($t_\star=650_{-500}^{+2250}$\,Myr), but was sufficient to rule out the THA membership scenario.

\item {\it HIP 90936  = HD 170773.}
F5V star with a broad, spatially resolved debris disk \citep{sepulveda19}.
Age determination is challenging, as typical for mid-F stars, and indeed a large scatter exists among the available measurements.
We determined EW(Li) $= 48\pm 5$ m\AA~ from the analysis of FEROS spectrum, which also confirms the fast rotation of the star (\vsini $=$ 50 km/s).
We adopted $800^{+400}_{-200}$\,Myr from isochrone fitting and the additional constraint from chromospheric activity.

\item {\it HIP 91043.}
G2V star, for which indirect methods yield an age $45^{+35}_{-25}$\,Myr \citep{vigan17}. The young age is further supported by the large spectroscopic and photometric variability \citep{willamo19,grandjean20}.
Isochrone fitting suggests an age $\sim 25$\,Myr. In order to account for the observed variability ($\sim 0.15$ mag in V band) and the time-dependent starspot fraction (0.03-0.15), we altered the photometry of the star by $\pm 0.1$ mag and used the SPOTS stellar models \citep{somers20}, allowing for a modeling of starspots, with spot fraction equal to 0 and 0.15. Accounting for all the uncertainties, we derived $t_\star=27 \pm 4$\,Myr.

\item {\it HIP 95793.}
Early type (A0IV), chemically peculiar star, and $\delta$ Scu pulsator.
Based on isochrone fitting, we estimated $t_\star=390_{-45}^{+40}$\,Myr.

\item {\it HIP 96334.}
Star with significant proper motion anomaly. Coupling the nondetection from SPHERE
data and the available HARPS RV timeseries \citep{grandjean23}, we inferred a mass most likely in the planetary regime for the astrometric perturber, as discussed in \citet{mesa22}.

\item {\it HIP 99137.}
F8V star, field star according to \banyan. The rotation period of 3.138 d \citep{fetherholf23} is intermediate between Hyades and Pleiades, arguing for an age around 250-300\,Myr. Similar values were obtained from the $\log{R'_{\text{HK}}}$ index, while the X-ray emission is closer to that of Hyades. The EW(Li) = $98.5 \pm 0.9$  m\AA from FEROS spectra and the isochrones are also compatible with this age. We adopted $t_\star=300_{-100}^{+200}$\,Myr.

\item {\it HIP 99945.}
G8/K0V star, field star according to \banyan. The kinematic parameters are within the kinematic box by \citet{montes01}, suggesting an age $\in [500, 1000]$\,Myr. 
The $\log{R'_{\text{HK}}}$ from CORALIE ($-4.38$ dex; Udry, priv. communication) is consistent with the first TESS period ($6.55$~d, from Mamajek's calibration, compared to an expected $P_{\text{rot}} \approx 7.0$ d). The value of $P_{\text{rot}}$ would indicate a $v_{\text{rot}} = 6.3$ km/s, considerably larger than the observed \vsini $= 3.0$  km/s; the required inclination ($28^\circ$), in any case, would not be so extreme as to rule out the scenario.
There are no archival spectra suitable for the measurement of  lithium.
The source is not revealed in X, but there are other cases of young objects missed by ROSAT. 
The most probable age from rotation/activity is therefore around 300\,Myr. 
The isochronal analysis is in tension with the indirect indicators: due to a redder color than the one expected at 300\,Myr ($\Delta (BP-RP) \approx 0.05$ mag, $\Delta (G-K) \approx 0.1$ mag), the star is compatible with either a pre-MS solution around 35\,Myr or a post-MS solution around 5\,Gyr. After ruling out unresolved multiplicity and a metallicity effect (the [Fe/H] of the star appears to be solar from the literature), we found that the discrepancy can be lifted assuming a mild reddening ($E(B-V) = 0.05$  mag) as in \citet{khramtsov21}. We adopted $t_\star=300_{-100}^{+500}$\,Myr, considering possible ambiguities in the derived rotation period.

\item {\it HIP 100787.}
A9III/IV star, field star according to \banyan. Lacking any indirect age indicator, we estimated $t_\star=430_{-60}^{+80}$\,Myr based on isochronal analysis.

\item {\it HIP 103460 = HD 199443 = HR 8018.}
A2/3III star with chemical peculiarities \citep{renson09}. 
It was removed from the F150 sample because of the RV variability observed by \citet{buscombe58} \footnote{RV dispersion: 20 km/s, mean error: 3 km/s,  six epochs over about 1000 days. No orbital solution available.}.
However, the analysis of DR3 data reveal no significant astrometric excess noise (RUWE$=$1.02) and no significant RV variability. The {\it Gaia-Hipparcos} PMA is also not significant ($\snrpma=0.85$).
We then considered the RV variability as unconfirmed and reincluded the star in the sample.
The {\it Hipparcos} and {\it Gaia} parallaxes are somewhat discrepant (3.8 $\sigma$ from the formal errorbars), possibly due to the bright magnitude ($V=5.89$  mag). The choice of the parallax has significant impact on the isochrone age, with {\it Gaia} indicating a more evolved object ($t_\star=730_{-30}^{+40}$\,Myr) than {\it Hipparcos} ($t_\star=500 \pm 200$\,Myr). We favored the {\it Gaia} solution due to the much smaller parallax uncertainty, but adjusted the lower age so as to account for the nominal parallax value from {\it Hipparcos}.

\item {\it HIP 104308.}
A5/6IV/V star with controversial membership to THA (supported by \citet{zuckerman04,malo13,bell15}; rejected by \citet{torres08,gagne18}). \banyan + Gaia DR3 kinematic data yield null membership probability.
Being an early type star, we relied on isochrone fitting, yielding $t_\star=725_{-50}^{+60}$\,Myr.

\item {\it HIP 105384.}
G7V star, whose rotation \citep[\prot $ = 5.036$~d][]{fetherholf23}, X-ray emission, and $\log{R'_{\text{HK}}}$ index are fully compatible with an age of 250\,Myr \citep{desidera15}. The detection of lithium, which would have been expected, was not reported by \citet{sacy}.

\item {\it HIP 105918.}
F8V star belonging to the field according to \banyan. The observed RV variability is in tension with the lack of astrometric hints of binarity; this is most likely explained by activity \citep{dalal21}, that would also be responsible for the signal previously attributed to a RV planet candidate \citet{butler17}.
The star has an IR excess \citep{ballering13}. Its age in the literature is generally assumed to be 900\,Myr following \citet{white07}. Our reanalysis argued for a younger age, although the picture is far from settled. Its kinematic is close to UMa, the \prot~(4.50 d) is compatible but highly sensitive to color uncertainties. The X-ray emission is slightly above Hyades, the $\log{R'_{\text{HK}}}$ (-4.56 dex) and the RV jitter are compatible with Hyades, while the EW(Li) (56$\pm$5  m\AA~ from SOPHIE spectra) lies below the Hyades locus (although some members have similar values). Accounting for all the uncertainties, we conservatively adopted $t_\star=500_{-200}^{+400}$\,Myr.

\item {\it HIP 108912.}
Moderately young and active solar type (G2V) star. The rotation period \citep[$P_{\text{rot}}=4.55$  d;][]{fetherholf23}, lithium, and activity  $\log R'_{HK}=-4.42$ dex \citep{henry96} consistently indicate an age of $300 \pm 100$\,Myr.

\item {\it HIP 111188.}
A1V star, classified as a field star by \banyan. Lacking any indirect age indicator, we estimated $t_\star=460 \pm 50$\,Myr based on isochronal analysis.

\item {\it HIP 114746.}
Field K2.5V star. 
The EW(Li) measured on a FEROS spectrum (77 m\AA), the chromospheric activity 
$\log R'_{HK}=-4.37$ dex \citep{gray06} and the rotation period $P_{\text{rot}} =7.877$ d \citep{fetherholf23} 
indicate an age similar to Group X and the TOI-1807 moving group (300\,Myr), while the star results younger than UMa and older than the Pleiades. We adopted therefore $t_\star=300 \pm 100$\,Myr.

\item {\it HIP 115527 = HD 220476 = NX Aqr.}
The star was considered as a probable member of the Octans-Near association
 by \cite{zuckerman13}.
The indirect age indicators are consistent with an age of $t_\star=150^{+150}_{-50}$ \,Myr.
The isochrone fitting yielded a consistent but more accurate age of $t_\star=150 \pm 50$\,Myr, that we then adopted for this target.

\item {\it HIP 116063.}
G1V star, classified as a field star by \banyan. Only loose constraints can be obtained from isochrone fitting, yielding a best-fit age of 150\,Myr with lower and upper ages equal to 40\,Myr and 600\,Myr.

\item {\it LP 876-10.}
Third component of the Fomalhaut (HIP 113368) system, host of a debris disk. The age was taken from \citet{mamajek12}. 

\item {\it TWA 14.}
Classified as a member of TWA in several papers \citep[e.g., ][]{zuckerman01b,malo13}; however it is not included in the members list in the recent analyses of TWA by \citet{luhman23} and \citet{miretroig25}.
The kinematic analysis with \banyan~ yielded a negligible membership probability to TWA, 13.5\% to LCC, 5.9\% to UCL, and 80.4\% to the field.
It is possible that the kinematic assignment is somewhat altered
by binarity, as the star has a significant excess noise (RUWE$=$1.727), probable RV variability from Gaia DR3, and additional RV measurement in the literature, although nominal RV errors may be underestimated for such a very active and fast rotating star.
The analysis with {\tt GaiaPMEX} indicated the presence of a companion ($\snrruwe = 6.9$), whose properties are however degenerate in the ($a$, mass) space. 
Various measurement of EW(Li) \citep{zuckerman01b,malo13,pecaut16} are well above those of BPIC, supporting an age younger than 15-20\,Myr.
For our isochronal analysis, we tested several values of the mass ratio in the range [0, 1]. The companion does not affect significantly the primary mass, that is well constrained to $0.70\pm0.02~\Msun$. Conversely, the age can range from 5 to 25\,Myr; imposing the constraint from lithium we can derived an upper limit on the mass ratio: $q \lesssim 0.7$. We adopted an age $13 \pm 7$\,Myr.
As the mass of the astrometric companion can fall into the substellar regime, we retained the star in the statistical sample.

\item {\it TYC 5736-0649-1.}
G6V star with extreme rotation  (\vsini~ $=$206 \kms, \citet{sacy};  \prot $=$0.25 d, \citet{desidera15}).
The possibility (mentioned in the above papers) that the object is a SB remains elusive after Gaia DR3, considering the increased errors due to \vsini~ and the lack of any significant astrometric scatter.

\item {\it TYC 6022-1079-1.}
K3V star slightly above main sequence locus, indicating a young age.
This is further supported by the rotation period \citep[0.989 d, ][]{kiraga12}
and EW(Li) \citep{sacy}, which is slightly lower than the Pleiades mean locus but within the distribution of the clusters members.
We adopted $t_\star=120_{-40}^{+80}$\,Myr from isochrone fitting.

\item {\it TYC 6069-1214-1 = BD-19 3018.}
K0V field star with age indicators intermediate between THA/COL/CAR and Pleiades/ABDO. We adopted the same age as in \citet{vigan17}.

\item {\it TYC 6243-0170-1.}
Young K2IV star with a very large EW(Li) \citep[420 m\AA;][]{sacy}, indicating an age younger than 20\,Myr.
\banyan~with Gaia DR3 parameters returned a 56.3\% membership probability for UCL (and 43.7\% for field).
The object appears to be on nearer side of the Sco-Cen association and have an age consistent with membership ($11 \pm 1$\,Myr).

\item {\it TYC 7476-0598-1.}
Young K2V star, proposed as a member of the Octans association by \citet{elliott16} but rejected from \banyan + Gaia DR3.
The large EW(Li) \citep[250 m\AA,][]{sacy} indicates an age on the order of 50\,Myr but is also compatible with THA on the young side and Pleiades and AB Dor on the old side within the intrinsic dispersion at their ages.
The rotation and activity are consistent with such young age. Coupling this with isochrones, compatible with any age $> 50$\,Myr, we adopted $t_\star=80_{-30}^{+70}$\,Myr.

\item {\it TYC 8174-1586-1.}
Classified as member of Carina association by \citet{sacy} but rejected by \banyan. Independently of the kinematic assignment, the very strong lithium line indicates an age younger than about 60-100\,Myr. We adopted $t_\star=40_{-15}^{+30}$\,Myr based on isochrone fitting.

\item {\it TYC 8175-118-1.}   
K6V star, classified as a field star by \banyan. Lacking any indirect age indicator, we estimated $t_\star=150_{-60}^{+350}$\,Myr based on isochronal analysis.

\item {\it TYC 8584-2682-1.}
The star was added to the sample as additional back-up (P4+, YMG member beyond 100 pc) after exhaustion of targets in the corresponding RA range.
Active G8V star classified as member of Carina by \citet{torres08},  of Columba by \citet{Elliott15}, and of the Platais 8 open cluster by \citet{qin23}.
The \banyan~ kinematic analysis with Gaia DR3 parameters yielded a null or very low membership probability to these groups, arguing instead for a field object.
The large EW(Li) from \citet{sacy} is just below the mean for THA, Columba, and Carina, well within the observed distribution of members of these associations, and clearly above the EW(Li) of Pleiades and ABDO members.
The fast rotation period from TESS \citep[1.92 d;][]{tu20,doyle20} is also compatible with a similar age. Isochrones suggested an even younger age of $23 \pm 2$\,Myr. We adopted the isochronal age as our baseline value, allowing for a maximum age of 50\,Myr.

\item {\it TYC 8634-1393-1.}
K5V star with a very wide companion, Gaia DR3 5343936233991962880 at 37.6" = 1560\,au. $\Delta G=$2.24 mag, SpT M3Ve \citep{skiff09} expected SpT M3/M3.5 from $BP-RP$ color.
Classified as a member of Carina association by \citet{torres08} and TWA by \citet{shan17}.
EW(Li) is both compatible with THA/Carina/Columba and Pleiades/AB Dor but not with a star as young as TWA.
The lithium non detection on the secondary \citep{lee22} is also consistent.
An isochronal age of 40\,Myr, consistent with indirect indicators, was derived for the primary. As regards the CMS, the picture is somewhat complicated by its clear unresolved binarity (RUWE$=$18.9), but a young age between 20 and 40\,Myr is possible assuming an equal-mass binary.

\item {\it TYC 8930-0601-1.}
New back-up target.
Member of Carina association in \citet{torres08}, whereas
\banyan~yields 86.8\% probability for field and 13.1\% for Platais 8.  \citet{vach24} argue instead for membership to the Theia92 group, with an age of 35\,Myr.
EW(Li), $P_{\text{rot}}$, $R_X$ are fully compatible with an age of about 40\,Myr, independently confirmed by isochrone fitting.

\end{description}

\section{Notes on individual objects: removed stars}
\label{a:notes_removed}

\subsection{Multiple stars}

\begin{description}

\item {\it 2MASS J02000918-8025009 = TYC 9357-1011-1.}
It was included in the sample as a THA member \citep[see, e.g.][]{kraus14}. However, \banyan~ yields null membership probability with Gaia DR3 and GaiaNSS kinematic values.
It is spatially resolved as a close binary (0.2-0.3") by \citet{tokovinin21,tokovinin22,mason23} with $q \approx1$.
In Gaia DR3 a large RUWE and astrometric acceleration (acceleration 9) are reported. 

Highly discrepant SpT in the literature (M1 vs.K3.7).
Photometric colors are fully compatible with K3.7 classification by \citet{kraus14}.
The position on CMD is not on pre-MS after correction of the magnitudes for binarity, further excluding the THA membership.
The EW(Li) by \citet{kraus14} and the rotation period by \citet{fetherholf23} are compatible with an age similar to the Pleiades and ABDO.

\item {\it 2MASS J05195513-0723399.}  
This star has a stellar companion at $0.5"=28$\,au, already described in \citet{bonavita22}, and is physically bound to another star at 15", 2MASS J05195412-0723359, which is itself a close visual binary.

\item {\it AF Hor = 2MASS J02414730-5259306.}
This star has a close (0.1") stellar companion, already described in \citet{bonavita22}, and is comoving with TYC 8491 656 1 (23" away).

\item {\it GSC 8057-00342.}
Identified as SB2 by \citet{flagg20}. Also astrometric binary according to {\tt GaiaPMEX}.

\item {\it GSC 8077-1788.}
Confirming the strong astrometric signature detected by RUWE (RUWE$=$2.5), this star is flagged as an astrometric binary in GaiaNSS.

\item {\it HD 100453.}
LCC star with an M-type companion at 1" = 120\,au companion detected by \citet{wagner15}.

\item {\it HD 106906.}
Planet-hosting system, belonging to LCC, composed of two F-type stars with a mass ratio close to unity \citep{bailey14,lagrange16}.

\item {\it HIP 14551.}
A3V star, member of Columba according to our \banyan~ analysis. F150 target. Both RUWE ($\snrruwe=30$) and PMa ($\snrpma=19$) clearly indicate the presence of a companion, whose minimum mass lies in the substellar regime ($M=60~\Mjup$). A mass of 0.091 $\Msol$ for this companion was inferred by \citet{gratton25} in a similar analysis. 
The star has a wide companion UCAC4 311-003056 at 59.34", with slightly discrepant astrometric parameters likely due to the close companion. 
If the X-ray source 1RXS J030750.4-275012 is associated with
the early type component ($\rho = 20.7"$), this would indicate
that the close companion is stellar. However, considering the typical errors on ROSAT position (about 30"), the association with the late-type companion ($\rho = 38.8"$) can not be ruled out. 
The additional evidence for the star being a binary is represented by the HARPS RV by \cite{lagrange09}: the radial velocity peak-to-valley amplitude observed between two groups of HARPS observations separated by $\sim 2$ yr is 3 km/s. Under the assumption of a circular orbit, the constraints on the \msini~ of the companion ruled out the small parameter space of the PMEX substellar solution at all separations.

\item {\it HIP 25434.}
This star has a close (0.06") stellar companion, already described in \citep{bonavita22}. Together with the spectroscopic binary HIP 25436 (separated by 12"), it forms a quadruple system.

\item {\it HIP 43299 = HD 75393.}
Astrometric ($\snrruwe = 68$, $\snrpma=835$) and spectroscopic binary without orbital solution.
Highly significant RV variable in \citet{nordstrom04} (7 measurements);
RV scatter of few km/s considering available literature measurements
\citep{nordstrom04,white07} and Gaia DR3. {\tt GaiaPMEX} constrains the secondary-to-primary mass ratio $q$ to be $> 0.3$; on the other hand, the $\chi^2$ of the isochrone fit for the system rapidly decreases assuming $q>0.7$. Based on our {\tt{MADYS}} analysis, we estimated $t_\star=250_{-150}^{+400}$\,Myr and $q \sim 0.5$ ($M_B \sim 0.6~M_\odot$).

\item {\it HIP 44722 = GJ 334 = BD-08 2582.}   
K7V/M0V star with a low mass companion (SpT M6), BD-08 2582B at $\rho = 8.35" \sim 121$\,au. 
There is some ambiguity in the literature about the photometric period. \citet{fetherholf23} report $P=0.615$ d for the primary while \citet{pass23} list the same period for the secondary, both from TESS data. However, the most likely explanation is that this photometric period belongs to the background RR Lyr variable ATO J136.6939-08.8023 at 24.1" from the primary, as a 0.615 d variability was observed at much higher spatial resolution
\citep[][ and Gaia DR2 RR Lyr catalog]{sesar17}.
A check on TESS data using 1-pixel aperture shows indeed the expected photometric variations for the RR Lyr star, and no clear photometric variations of the other targets with the proposed periodicity. 
The age of 80\,Myr for the primary, suggested by isochrone fitting, is not compatible with the robust $>$ 1\,Gyr constraint coming from the CMS. There is no indication from the literature or from {\tt GaiaPMEX} that the observed over-luminosity of the primary is due to unresolved multiplicity. The tension can be alleviated by assuming a super-solar metallicity ([Fe/H] $\gtrsim 0.15$), as suggested by some works \citep{soubiran16,hojjatpanah19}, leading to a broad age constraint $t_\star=1000_{-800}^{+4000}$\,Gyr.

\item {\it HIP 49366 = HD 87424 = V417 Hya.}
Wide binary system with HD 87424B at 9.7"$\sim$225\,au. The spectral type of the secondary, derived from the $BP-RP$ color and \cite{pecaut13} main-sequence relationships\footnote{Updated version at \url{https://www.pas.rochester.edu/~emamajek/EEM_dwarf_UBVIJHK_colors_Teff.txt}.}, is M5/M5.5, which would correspond to about 0.14 $M_\odot$. The large proper motion difference ($\Delta \mu_\alpha \cos{\delta} = 12$ mas/yr, $\Delta \mu_\delta = 7$ mas/yr) between the components is compatible with orbital motion, assuming the orbit is seen face-on.
Ambiguous rotation period from \citet{strassmeier00} and \citet{fetherholf23}, differing by about a factor of two.
The M-star companion puts a tight age constraint: $t_\star>300$\,Myr. We adopted $t_\star=500_{-200}^{+500}$\,Myr.

\item {\it HIP 55899.}
A0V star, showing a comoving object -- CD-39 7118B, hereafter B -- at 22" = 2900\,au. Both objects have a large RUWE: 5.6 for A, 7.6 for B. A significant PMa ($\snrpma=15$) is found for A, further strengthening the idea of a high multiplicity system. 
Indicated as a Sco-Cen star in pre-{\it Gaia} studies \citep{dezeeuw99,rizzuto11}, but a very low membership probability is found by \banyan~ (2\%). This might be a consequence of the astrometric perturbation from the unseen companion.
Bright source in ROSAT. B shows a very large  EW(Li) \citep[662  m\AA;][]{pecaut16}, from which a very young age ($<$20-30\,Myr), compatible with Sco-Cen, was inferred.
Isochronal analysis was complicated by the poorly unconstrained $q$. On the one hand, {\tt GaiaPMEX} rejected any substellar companion (more precisely, any source with $q<0.15$); on the other hand, values $q>0.6$ were ruled out due to a quick degeneration of the best-fit $\chi^2$. Within the range $0.15 < q < 0.6$, any age between 10 and 30\,Myr is possible. We adopted $t_\star=20 \pm 10$\,Myr, and derive an upper limit of $1.2~M_\odot$ for the unseen companion of A.

\item {\it HIP 57013 = HD 101615.}
A0V star with small projected rotational velocity, bona fide member of Argus according to \banyan. The star has a wide early M companion (2MASS J11411975-4305522, hereafter HD 101615B) at 8.4", whose physical association is confirmed by {\it Gaia}. The age of HD 101615B from isochrones is younger than the one of Argus ($24\pm4$\,Myr), but this might be due to the fact that B itself is likely an unresolved binary ($\ruwe=2.2$).
Analysis of HARPS archive data (30 spectra spanning about 1100 days) reveals that the star is a spectroscopic binary with period of 2.208 days and RV semi-amplitude of 18.9 km/s. This corresponds to $m\sin{i}= 0.19~\Msol$. Assuming that the system is tidally locked (\prot $=P_{\text{orb}}$) we inferred from the observed \vsini that the system is viewed nearly face-on ($i=26^\circ$). In case of co-planarity, we expect a true mass for the companion of 0.43 M$_\odot$.

\item {\it HIP 58146.}
Confirming the strong astrometric  signatures ($\snrruwe =7.0$, $\snrpma = 7.1$), this star is flagged as an SB1 in GaiaNSS.

\item {\it HIP 61049.}
Confirming the strong astrometric  signatures ($\snrruwe =52$, $\snrpma = 3.6$), this star is flagged as an SB1 in GaiaNSS.

\item {\it HIP 64322.}
This star is triple system, having a moderately wide
companion at 2.3" = 235\,au projected separation \citep{bonavita22}, and a RV trend with curvature over a baseline of 680 days, indicating the presence of an inner companion \citep{grandjean23} which {\tt GaiaPMEX} ($\snrpma=108$) constrains to have $M>0.1~\Msol$.

\item {\it HIP 66722.}
A0V member of UCL. Both astrometric signatures ($\snrruwe = 9.9$, $\snrpma=6.1$) point toward the existence of a companion, that must be stellar according to {\tt GaiaPMEX}. A high-mass companion is needed to reconcile the observed photometry with the group age.

\item {\it HIP 70833.}
This star is a triple composed of a spectroscopic binary and a K-type star located 2.9" away \citep{bonavita22}. The secondary is detected by TESS (\citetalias{desidera21}).

\item {\it HIP 72192.}
A0V star, bona fide member of UCL. Both the RUWE ($\snrruwe=12$) and the PMa ($\snrpma=92$) point toward the existence of a companion, that must be stellar according to {\tt GaiaPMEX}. Indeed, a high-mass companion ($q \in$ 0.3-0.6) is needed to reconcile the observed photometry with the group age.

\item {\it HIP 72622.}
Naked-eye A3IV/V star belonging to the field, it is a spectroscopic binary with mass ratio $q=0.92$ \citep{fuhrmann14,waisberg23} and a semi-major axis $a = 0.52$\,au. The system is gravitationally bound to HIP 72603 (231"), itself a spectroscopic binary \citep{beuzit04}, making it a quadruple system. We estimated $t_\star=700_{-200}^{+300}$\,Myr based on isochrone fitting.

\item {\it HIP 88694 = HD 165185.}
Star classified as a member of UMa group by \citet{montes01}, but field object according to \banyan. 
It has a visual companion at 12.3" = 210\,au. The B component (2MASS J18062369-3601237) has SpT M0Ve.
Independently of the kinematic membership, the indirect age indicators (rotation period, lithium, $R_X$, and $\log{R'_{\text{HK}}}$) are fully consistent with the UMa age range. The CMS does not provide any additional constraint beyond $t_\star > 200$\,Myr.

\item {\it HIP 101483 = HD 195943 = $\eta$ Del.}
A3IV star, belonging to the field according to our \banyan~ analysis. Both the RUWE and the PMa are highly significant ($\snrruwe=29$, $\snrpma=88$), but there is no astrometric orbital solution in Gaia DR3 or direct detection in our SPHERE images. {\tt GaiaPMEX} constrains the mass ratio to be $0.2<q<0.9$, firmly in the stellar regime. 
The binarity was confirmed by the SB orbital solution by \cite{grandjean21} (period of $1363$~d, eccentricity 0.09, mass $0.54~\Msol$). Isochronal analysis yielded $t_\star=100_{-70}^{+300}$\,Myr.

\item {\it HIP 102626 = HD 197890 = BO Mic.}
After various but inconclusive evidences for binarity, summarized in \citetalias{desidera21}, the GaiaNSS astrometric solution conclusively shows the presence of a companion. This is the only star in the F150 sample with a GaiaNSS solution.

\item {\it HIP 105140 = HD  202627 = $\epsilon$ Mic.}
Chemically-peculiar A0V star.
Significant astrometric  signatures ($\snrruwe =6.7$, $\snrpma =3.0$), indicating a rather massive companion. Indeed, {\tt GaiaPMEX} constrains its mass to be stellar ($M > 90 M_J$).
The small RV error in {\it Gaia} and the moderately low scatter of RV measurements in the literature indicate an upper limit of about 1-2 km/s on the RV semi-amplitude.
The CMD position above the MS allowed us to derive a tight determination of stellar age ($t_\star=650 \pm 50$\,Myr).

\item {\it HIP 111449.}
F7V star, considered in earlier works \citep{lafreniere07} as a member of the Hercules-Lyra moving group -- whose existence, however, is disputed \citep{riedel17}. Our \banyan~ analysis classifies it as a field star. A CMS, Gaia DR3 6628926642897199744, exists at 6" $= 140$\,au; its relative proper motion and RV compared to HIP 111449 are compatible, within the uncertainties, with orbital motion. However, its faintness ($G=11$) argues for a substellar mass ($60-80~M_{\text{Jup}}$) impeding any independent age constraint. As in \citet{vigan17}, we adopted a broad $t_\star=250_{-50}^{+750}$\,Myr.

\item {\it HIP 113201.}
This star has a low-mass stellar companion at $0.16"=4$\,au, already described in \citet{bonavita22}. \banyan~ identifies it as a member of BPIC; however, the position of the star in the CMD is incompatible with such a young age. We retained the age estimate provided in \citet{bonavita22}.

\item {\it HIP 114952 = HD 219592.}
F5V star, classified as a field star by \banyan. Both the RUWE and the PMa are highly significant ($\snrruwe)=106$, $\snrpma)=6.5$), with {\tt GaiaPMEX} constraining the unseen secondary to be in the stellar regime. Additional hints for binarity come from the astrometric orbital solution in GaiaNSS, a discrepant parallax between {\it Gaia} and {\it Hipparcos}, and a clear variability ($> 2$ km/s seen in 11 spectra in SOPHIE archive (epoch 2008-2010). 
The star was selected as young because of the value of $\log{R'_{\text{HK}}} = -4.21$ dex by \cite{gray06}, but it is likely that this measurement is spurious, as found for other F type stars from the same study (see \citealt{desidera15} for details). The true age of the star is not well constrained, due to the unconstrained secondary mass; if the secondary is as massive as the primary, $t_\star \in [50, 400]$\,Myr; conversely, a secondary that does not significantly contribute to the total flux implies a system age around 2\,Gyr.

\item {\it HIP 118121 = HD 224392 = $\eta$ Tuc.}
A1V star, classified as member of THA in several papers \citep{zuckerman04,malo13}. The result of our \banyan~analysis depends on the adopted RV: without RV there is a 99.9\% membership probability to the association, with a predicted RV for membership of 6.3 km/s. However, the RV is discrepant in the literature (-32.5 km/s in GCRV, 0.57 km/s in \citealt{Malaroda06}), and is also an RV variable \citep{Borgniet17}, with peak-to-valley variations of 8 km/s. The star, having a strong astrometric signal from RUWE ($\snrruwe)=20$), is listed in the GaiaNSS catalog (binary period $413$~d). This companion was directly detected with the PIONIER interferometer at VLTI \citep{marion14}, and naturally explains the observed X-ray emission.
The two objects listed in the WDS and detected at 13'' by \citet{ehrenreich10} are background stars, as confirmed by {\it Gaia}.

\item {\it TYC 7722-0207-1 = HD 296790.}
Fast-rotating K0 dwarf, with age indicators leading to an age similar to the Pleiades  \citep{desidera15}. \citet{traven20} identified the star as an SB2 with a mass ratio close to unity. The unresolved multiplicity allowed us to reconcile the CMD position with the indirect indicators.

\item {\it TYC 8137-2609-1 = CD-46 3212.}
Indicated as a K0V star in Simbad, it is a field star according to \banyan. It has a strong lithium in SACY ( EW(Li)$=300$ m\AA) and is also X-ray bright, implying an age around 20\,Myr. However, the parallax from {\it Gaia} ($2.43\pm0.01$  mas) is incompatible with a dwarf star.
{\it Gaia}'s RV differs by 25 km/s from the SACY one, and there is a large $\Delta \mu$ signature between Tycho2 and Gaia DR3 ($\Delta \mu = 8.6 \pm 3.2$ mas/yr), confirming that the star is a spectroscopic binary.
On the other hand, astrometric indicators in Gaia DR3 do not support binarity. Photometric periods of 0.843 d and 1.224 d have been reported in the literature \citep[][respectively]{kiraga12,richards12}. One of the two is the alias of the true period. Our analysis of ASAS time series favors $P=1.224$ d as the true one. The rotation period coupled with the \vsini$=160$ km/s (SACY) indicate a radius $R_\star > 2.6~R_\odot$, which implies $L_\star >  3.8~L_\odot$ and $d > 370$ pc, a value compatible with the {\it Gaia} parallax. Isochrone fitting yields $t_\star=3.7 \pm 1.2$\,Gyr, $M_\star=1.35 \pm 0.05~M_{\odot}$. We conclude that the system is an evolved star. The very fast rotation and X-ray emission can be explained if the system is a tidally-locked system.
The high lithium content is also observed in other RS CvN systems \citep{pallavicini92b}. The tidally-locked companion is too close to produce detectable $\Delta \mu$ signatures, possibly hinting at a tertiary component. However, this possible additional component is not seen in SPHERE images.

\item {\it TYC 8491-1376-1.}
K4IV star, which \banyan~ labels as a field star. The extremely large value of the RUWE (3.27, corresponding to $\snrruwe =21$) argues for the presence of an unresolved stellar companion. An age of 30\,Myr was inferred by \citet{zerjal17} based on the CaII infrared triplet. The age was confirmed by isochronal analysis, assuming that the system is composed by two stars with $q \sim 1$.
The star is listed in the GaiaNSS catalog, with a binary period of 118 d. An equal-mass binary with $a \approx 0.5$\,au is fully consistent with  {\tt GaiaPMEX} maps.

\end{description}

\subsection{Old stars}

\begin{description}

\item {\it HIP 39826 = BD+21 1764.}
K7V star with an M-type CMS, BD+21 1764B, at $10" = 180$\,au), which is itself a tight triple; SB2 from \citep{shkolnik10} and close visual \citep[$\rho$ 0.33"$-$0.97" from 2012 to 2021, ][]{cortes17,tokovinin22}. Both chromospheric activity ($\log{R'_{\text{HK}}}=-4.80$ dex \citep{brown22} and isochrones indicate an age between 2 and 4\,Gyr.
The prominent X-ray emission might be explained by the contribution of the four components and possibly by tidal locking of the SB2 (the orbital period is not known). Therefore, we did not consider it for age determination. 

\item {\it HIP 73666.}  
F3IV star \citep{houk82}, classified as a Sco-Cen member in early studies \citep[e.g.][]{dezeeuw99} but rejected by \banyan~ ($p<1\%$). A significant reddening $E(B-V)=0.12$-0.15  mag is inferred by \citet{casagrande11} and \citet{pecaut12}, and is confirmed in the recent reanalysis by \citep{paunzen24} based on several photometric indices. Assuming this reddening, the isochronal analysis yields either $t_\star=5 \pm 1$\,Myr (pre-MS solution) or $t_\star=1300 \pm 100$\,Myr (post-MS solution).
No evidence was found for the existence of a CMS group related to the star.
We considered therefore the post-MS solution as being more likely.

\item {\it HIP 77015.}   
The classification of this star as a Sco-Cen member (UCL) by \citet{dezeeuw99} was later rejected by \citet{mamajek02}. Indeed, it is found to be a field object in our kinematic analysis.
Discrepant spectral classification (G3V in {\it Hipparcos}, {G0.5V} in \citealt{mamajek02}), with photometric colors supporting the latter classification, possibly even a slightly earlier one (F9.5/G0). 
The analysis of the HARPS spectra by \citet{grandjean23} shows low RV scatter (2.4 m/s for 15 spectra over 667 days) and low activity. This is further supported by the X-ray non detection by ROSAT.
The EW(Li) from our analysis of HARPS spectra (EW(Li) = 77.1$\pm$4.4 m\AA) is intermediate between Hyades and NGC 752 members of similar colors, suggesting $t_*\sim$ 1\,Gyr.
Isochrones suggest an even older age ($t_\star=4 \pm 1$\,Gyr). We concluded that the star is an old interloper.

\item {\it HIP 92283 = HD 174080.}   
K3/K4 star from photometric colors \citep{martinez10}, in some tension with the reported K0.
The star appears to be older than the bulk of SHINE stars.
Kinematic space velocities are outside the box typical of young stars \citep{montes01}.
A moderate chromospheric and coronal activity is observed, compatible but lower than the mean value of Hyades members of similar colors.
The nondetection of lithium adds only limited constraints considering the late spectral type.
There is no measurement of rotation period; the ages expected based on the $\log{R'_{\text{HK}}}$ values by \citet{duncan91} and \citet{gomesdasilva21} are 1.1 and 1.7\,Gyr.

\item {\it HIP 96085 = HD 183870.}
K2V star forming a very wide binary system ($\rho = 208.6" \sim 3700$\,au) with the M5 star HD 183870B.
A fast rotation period (2.32 d) and $H_{\alpha}$ emission were measured on the secondary by \citet{pass22}, with indication of an age older then 1\,Gyr, while the rotation period of the primary remains undetected.
From the S index measurement by \cite{wright04}, an expected rotation period
of 18.7d was inferred, corresponding to a gyro-age of 1.3\,Gyr.
The X-ray and lithium non detection are compatible with a moderately old star.
We then adopted $1300 \pm 300$\,Myr.

\end{description}

\subsection{Accreting or gas-rich}

\begin{description}

\item {\it HIP 82323 = AS 209.}  
The star is very young, as shown by the presence of a protoplanetary disk imaged by ALMA \citep{andrews18}. Our analysis confirms an age $<$ 2\,Myr.
The reddening is uncertain and likely very large ($E(B-V) \sim 1$ mag). According to {\tt GaiaPMEX} ($\snrruwe)=3.7$), an astrometric companion with unknown $q$ -- possibly extending to planetary masses -- exists; however, for such young stars, spurious astrometric signals may arise due to accretion \citep[see, e.g.,][]{lagrange25}.

\end{description}

\section{Tables}

\onecolumn

\begin{landscape}
\begin{longtable}{llccccccl}
\caption{Main astrometric, photometric, and spectral parameters for the F400 sample.}
\label{t:basic_astro_data} \\
\hline\hline 
Name & Gaia DR3 & $\alpha$ & $\delta$ & $\varpi$ & $G$ & $H$ & SpT & Disk \\
 & & (hms) & (dms) & (mas) & (mag) & (mag) & & \\
\hline
\endfirsthead
\caption{continued.}\\
\hline\hline
Name & Gaia DR3 & $\alpha$ & $\delta$ & $\varpi$ & $G$ & $H$ & SpT & Disk \\
 & & (hms) & (dms) & (mas) & (mag) & (mag) & & \\

\hline

\hline
\endhead

\hline
\endfoot

\hline 
\endlastfoot

HIP 490 & Gaia DR3 4995853014646189696 & 0 05 52.5447 & -41 45 11.044 & $25.75 \pm 0.02$ & 7.40 & 6.19 & G0V & IREX, RES $^b$ \\
HIP 682 & Gaia DR3 2745647973856851712 & 0 08 25.7454 & 6 37 00.489 & $26.04 \pm 0.03$ & 7.45 & 6.15 & G2V & IREX, RES $^{a,b}$ \\
HIP 1113 & Gaia DR3 4688292536884945536 & 0 13 53.0108 & -74 41 17.85 & $21.80 \pm 0.01$ & 8.65 & 7.09 & G8V & no \\
2MASS J00152752-6414545 & Gaia DR3 4900372657545000192 & 0 15 27.5152 & -64 14 54.769 & $20.97 \pm 0.01$ & 11.93 & 8.69 & M1.8 & no \\
2MASS J00172353-6645124 & Gaia DR3 4707563810327288192 & 0 17 23.5412 & -66 45 12.332 & $27.16 \pm 0.02$ & 11.34 & 7.93 & M3.0Ve & no \\
HIP 1481 & Gaia DR3 4901229043960053248 & 0 18 26.1235 & -63 28 38.98 & $23.36 \pm 0.02$ & 7.34 & 6.25 & F8V & IREX \\
HIP 1993 & Gaia DR3 4902014095262270336 & 0 25 14.6618 & -61 30 48.252 & $22.65 \pm 0.01$ & 10.69 & 7.94 & M0Ve & no \\
HIP 2578 & Gaia DR3 4900916679578514048 & 0 32 43.9053 & -63 01 53.402 & $21.68 \pm 0.07$ & 5.06 & 5.16 & A0V & IREX \\
\end{longtable} 	
\tablefoot{For disks, IREX = infrared excess from \citet{chen14}; RES: resolved disk ($^a$: catalog of circumstellar disks; $^b$: \citealt{engler25}; $^c$: \citealt{ingleby11}).}
\end{landscape} 

\def \snr{\text{S/N}}
\def \ruwe{\text{RUWE}}
\def \pma{\text{PMa}}

\begin{landscape}
\begin{longtable}{lccccccccccc}
\caption{Kinematic properties of F400 stars from Gaia DR3, membership to YMG from BANYAN $\Sigma$, and astrometric indicators of binarity.}
\label{t:basic_kin_data} \\
\hline\hline 
Name & $\mu_\alpha \cos{\delta}$ & $\mu_\delta$ & Ref. & $\gamma$ & YMG & RUWE & $\snr(\ruwe)$ & $\snr(\pma)$ & $M_{2, \text{min}}$ \\
 & (mas yr$^{-1}$) & (mas yr$^{-1}$) & & (km s$^{-1}$) & & & & & ($M_{\text{Jup}}$) \\
\hline
\endfirsthead
\caption{continued.}\\
\hline\hline
Name & $\mu_\alpha \cos{\delta}$ & $\mu_\delta$ & Ref. & $\gamma$ & YMG & RUWE & $\snr(\ruwe)$ & $\snr(\pma)$ & $M_{2, \text{min}}$ \\
 & (mas yr$^{-1}$) & (mas yr$^{-1}$) & & (km s$^{-1}$) & & & & & ($M_{\text{Jup}}$) \\
           
\hline

\hline
\endhead

\hline
\endfoot

\hline 
\endlastfoot

HIP 490 & $97.94 \pm 0.01$ & $-76.51 \pm 0.01$ & G3H & $2.05 \pm 0.16$ & THA & 0.91 & 0.31 & 1.07 & ... \\
HIP 682 & $87.47 \pm 0.03$ & $-1.83 \pm 0.01$ & G3H & $0.93 \pm 0.14$ & Field & 1.03 & 1.18 & 0.62 & ... \\
HIP 1113 & $82.71 \pm 0.03$ & $-49.14 \pm 0.03$ & G3H & $9.09 \pm 0.19$ & THA & 0.83 & 0.19 & 0.29 & ... \\
2MASS J00152752-6414545 & $79.94 \pm 0.02$ & $-54.33 \pm 0.02$ & G3 & $5.97 \pm 24.00$ & THA & 1.10 & 1.26 & ... & ... \\
2MASS J00172353-6645124 & $102.89 \pm 0.02$ & $-16.83 \pm 0.02$ & G3 & $9.42 \pm 0.53$ & BPIC & 0.95 & 0.37 & ... & ... \\
HIP 1481 & $90.14 \pm 0.02$ & $-59.21 \pm 0.02$ & G3H & $6.53 \pm 0.16$ & THA & 0.99 & 0.78 & 1.57 & ... \\
HIP 1993 & $88.00 \pm 0.07$ & $-56.24 \pm 0.08$ & G3H & $6.21 \pm 0.38$ & THA & 1.02 & 0.78 & 0.77 & ... \\
HIP 2578 & $86.49 \pm 0.01$ & $-50.11 \pm 0.01$ & G3H & $3.60 \pm 0.25$ & THA & 1.03 & 0.76 & 0.09 & ... \\
\end{longtable} 	
\tablefoot{References for proper motions: G3 = Gaia DR3, G2 = Gaia DR2, H = Hipparcos, G3H = long-term Gaia-Hipparcos proper motion.}
\end{landscape} 

\begin{landscape}
\begin{longtable}{lccccccll}
\caption{Indirect age indicators.}
\label{t:indirect_age_indicators} \\
\hline\hline 

Name & $\log{L_{\rm X}}$ & $R_{\rm X}$ & $\log{R'_{\rm HK}}$ & $P_{\rm rot}$ & $v \sin{i}$ & EW(Li) & Ref. & Remarks \\
 & ($10^{-7}$ W) & & (dex) & (d) & (km s$^{-1}$)& (m{\AA}) & & \\
\hline
\endfirsthead
\caption{continued.}\\
\hline\hline
Name & $\log{L_{\rm X}}$ & $R_{\rm X}$ & $\log{R'_{\rm HK}}$ & $P_{\rm rot}$ & $v \sin{i}$ & EW(Li) & Ref. & Remarks \\
           
\hline

\hline
\endhead

\hline
\endfoot

\hline 
\endlastfoot

HIP 490 & ... & ... & ... & ... & ... & ... & ... & ... \\
HIP 682 & ... & ... & ... & ... & ... & ... & ... & ... \\
HIP 1113 & ... & ... & ... & ... & ... & ... & ... & ... \\
2MASS J00152752-6414545 & ... & ... & ... & ... & ... & ... & ... & ... \\
2MASS J00172353-6645124 & ... & ... & ... & ... & ... & ... & ... & ... \\
HIP 1481 & ... & ... & ... & ... & ... & ... & ... & ... \\
HIP 1993 & ... & ... & ... & ... & ... & ... & ... & ... \\
HIP 2578 & ... & ... & ... & ... & ... & ... & ... & ... \\
\hline

\end{longtable} 	
\tablefoot{Indicators: X-ray luminosity ($\log{L_{\rm X}}$), fractional X-ray luminosity ($R_{\rm X}$), chromospheric R'HK index ($\log{R'_{\rm HK}}$), rotation period ($P_{\rm rot}$), rotational velocity ($v \sin{i}$), equivalent width of the 6708 {\AA} lithium line (EW(Li)). References: A17 = \citet{astudillo17}; B19 = \citet{bowler19}; B22 = \citet{brown22}; C24 = \citet{colman24}; D15 = \citet{desidera15}; F22 = \citet{fuhrmeister22}; F23 = \citet{fetherholf23}; G20 = \citet{gondoin20}; H19 = \citet{hojjatpanah19}; K14 = \citet{kraus14}; L09 = \citet{lopez09}; M13 = \citet{malo13}; M16 = \citet{messina16}; P16 = \citet{pecaut16}; P24 = \citet{perdelwitz24}; R12 = \citet{richards12}; R18 = \citet{reiners18}; S00 = \citet{strassmeier00}; S07 = \citet{schroeder07}; S24 = \citet{shan24}; T06 = \citet{torres06}; T22 = \citet{tranin22}; TW = this work; W04 = \citet{wright04}. n.d. indicates a non detection.}
\end{landscape} 

\def \Msol{M_\odot}             

\begin{landscape}
\begin{longtable}{lccccccccl}
\caption{Final ages and masses for all SHINE stars and inclusion in the final statistical sample.}
\label{t:ages_masses} \\
\hline\hline 
Name & $t^*_\text{opt}$ & $t^*_\text{min}$ & $t^*_\text{max}$ & Method & $M^*_\text{opt}$ & $M^*_\text{min}$ & $M^*_\text{max}$ & Stat. & Remarks \\
 & (Myr) & (Myr) & (Myr) & & ($\Msol$) & ($\Msol$) & ($\Msol$) & & \\
\hline
\endfirsthead
\caption{continued.}\\
\hline\hline
Name & $t^*_\text{opt}$ & $t^*_\text{min}$ & $t^*_\text{max}$ & Method & $M^*_\text{opt}$ & $M^*_\text{min}$ & $M^*_\text{max}$ & Stat. & Remarks \\
 & (Myr) & (Myr) & (Myr) & & ($\Msol$) & ($\Msol$) & ($\Msol$) & & \\
           
\hline

\hline
\endhead

\hline
\endfoot

\hline 
\endlastfoot

HIP 490 & 37 & 26 & 48 & YMG & 1.11 & 1.04 & 1.12 & T & ... \\
HIP 682 & 150 & 100 & 200 & IM & 1.08 & 1.07 & 1.09 & T & ... \\
HIP 1113 & 37 & 26 & 48 & YMG & 0.89 & 0.88 & 0.93 & T & ... \\
2MASS J00152752-6414545 & 37 & 26 & 48 & YMG & 0.57 & 0.55 & 0.58 & T & ... \\
2MASS J00172353-6645124 & 21 & 17 & 25 & YMG & 0.53 & 0.50 & 0.55 & T & ... \\
HIP 1481 & 37 & 26 & 48 & YMG & 1.16 & 1.15 & 1.18 & T & ... \\
HIP 1993 & 37 & 26 & 48 & YMG & 0.70 & 0.68 & 0.71 & T & ... \\
HIP 2578 & 37 & 26 & 48 & YMG & 2.04 & 2.03 & 2.05 & T & ... \\
\hline

\end{longtable} 	
\tablefoot{Stat.: boolean indicating whether the star was retained (T) or not (F) in the statistical sample. Remarks: reasons for exclusion from the statistical sample. Method: age determination method (ACT = activity; ISO = isochrones; Li = lithium; YMG = YMG membership; ROT = rotation; IM = multiple indirect methods; SC-c: Sco-Cen, cms; SC-i: Sco-Cen, individual; SC-s: Sco-Cen, subgroup; SC-cis, SC-ci, SC-is indicate the combination of multiple methods). Remarks: IC = companion resolved in DI or interferometry; OS = star older than 1 Gyr; NSS = GaiaNSS binary; PM = GaiaPMEX binary; OA = other astrometric binary; SB = spectroscopic binary; GR = gas-rich disk.}
\end{landscape}

\begin{landscape}
\begin{longtable}{lcclllcccccl}
\caption{Known stellar companions to the stars in our sample. For each companion, we compute its best-fit mass ($M_2$), angular separation ($\rho_2$), and projected separation ($d_2$); if $\rho_2$ is not known, we report the upper limit $\rho_{2,\rm UL}$ coming from the combination of all available data and the corresponding $d_{2,\rm UL}$.}
\label{t:stellar_companions} \\
\hline\hline 
Name & M & Method & N1 & Ref. & Companion name & $M_2$ & $\rho_2$ & $d_2$ & $\rho_{2,\rm UL}$ & $d_{2,\rm UL}$ & N2 \\
 & & & & & & $(\Msol)$ & (arcsec) & (au) & (arcsec) & (au) & \\
\hline
\endfirsthead
\caption{continued.}\\
\hline\hline
Name & M & Method & N1 & Ref. & Companion name & $M_2$ & $\rho_2$ & $d_2$ & $\rho_{2,\rm UL}$ & $d_{2,\rm UL}$ & N2 \\
 & & & & & & $(\Msol)$ & (arcsec) & (au) & (arcsec) & (au) & \\
\hline

\hline
\endhead

\hline
\endfoot

\hline 
\endlastfoot

2MASS J02000918-8025009 & 2 & NSS, IC & ... & a9, T22 & ... & 0.74 & 0.024 & 1.8 & ... & ... & ... \\
2MASS J05195513-0723399 & 2 & IC & ... & B22 & ... & 0.13 & 0.49 & 26 & ... & ... & ... \\
AF Hor & 2 & IC & ... & B22 & ... & 0.55 & 0.050 & 2.2 & ... & ... & ... \\
GSC 08057-00342 & 1 & RV & SB2 & F20 & ... & 0.43 & ... & ... & 0.10 & 4.4 & ... \\
GSC 08584-01898 & 2 & IC & ... & B22 & ... & 0.13 & 0.26 & 35 & ... & ... & ... \\
GSC 8077-1788 & 1 & NSS & ... & orb & ... & ... & ... & ... & 0.010 & 0.84 & ... \\
HD 100453 & 2 & IC & ... & W15 & ... & 0.30 & 1.1 & 110 & ... & ... & ... \\
HD 106906 & 1 & RV & SB2 & L16 & ... & 1.4 & ... & ... & 0.010 & 1.0 & ... \\
HD 147553A & 3 & R & ... & G & HD 147553B  & ... & 6.2 & 860 & ... & ... & ... \\
HD 25284 & 2 & IC & ... & B22 & ... & 0.46 & 0.073 & 3.7 & ... & ... & ... \\
HD 61606B & 3 & R & ... & G & HIP 37349 & 0.81 & 58 & 820 & ... & ... & ... \\
HIP 101483 & 1 & PM & ... & TW & ... & ... & ... & ... & 0.10 & 5.5 & ... \\
HIP 102626 & 1 & PM, NSS & ... & ots & ... & ... & 0.040 & 2.0 & ... & ... & ... \\
HIP 105140 & 1 & PM & ... & TW & ... & ... & ... & ... & 0.10 & 5.2 & ... \\
HIP 107948 & 2 & IC & ... & B22 & ... & 0.45 & 0.19 & 5.8 & ... & ... & ... \\
HIP 107948 & 2 & IC & ... & B22 & ... & 0.22 & 0.66 & 20 & ... & ... & ... \\
HIP 109285 & 2 & PM, IC & ... & B22 & ... & 0.94 & 0.060 & 2.2 & ... & ... & ... \\
HIP 109427 & 2 & IC & ... & B22 & ... & 0.21 & 0.22 & 6.1 & ... & ... & ... \\
HIP 111449 & 3 & R & ... & G & Gaia DR3 6628926642897199744 & ... & 6.0 & 140 & ... & ... & ... \\
HIP 112581 & 2 & IC & ... & B22b & ... & 0.14 & 0.74 & 28 & ... & ... & ... \\
HIP 113201 & 2 & IC & ... & B22 & ... & 0.10 & 0.16 & 3.8 & ... & ... & ... \\
HIP 113283 & 3 & R & ... & G & Fomalhaut A & 1.9 & 7000 & 53000 & ... & ... & ... \\
HIP 113283 & 3 & R & ... & G & Fomalhaut C & 0.23 & 27000 & 210000 & ... & ... & ... \\
HIP 113368 & 3 & R & ... & G & Fomalhaut C & 0.23 & 20000 & 150000 & ... & ... & ... \\
HIP 113368 & 3 & R & ... & G & Fomalhaut B & 0.74 & 7000 & 54000 & ... & ... & ... \\
HIP 114952 & 1 & NSS, PM, RV & ... & orb & ... & ... & 0.10 & 4.8 & ... & ... & ... \\
HIP 118121 & 1 & PM, IC, NSS & ... & M14 & ... & 1.4 & 0.034 & 1.7 & ... & ... & ... \\
HIP 14551 & 1 & RV, PM & ... & TW, L09 & ... & ... & ... & ... & 0.060 & 3.2 & ... \\
HIP 17157 & 2 & IC & ... & B22 & ... & 0.28 & 1.8 & 45 & ... & ... & ... \\
HIP 17157 & 2 & IC & ... & B22 & ... & 0.19 & 1.4 & 36 & ... & ... & ... \\
HIP 17797 & 2 & PM, IC & ... & B22 & ... & 0.55 & 0.12 & 6.4 & ... & ... & ... \\
HIP 18714 & 2 & IC & ... & B22 & ... & 0.25 & 0.044 & 2.3 & ... & ... & ... \\
HIP 19183 & 2 & IC & ... & B22 & ... & 0.35 & 4.2 & 240 & ... & ... & ... \\
HIP 25434 & 2 & IC & ... & B22 & ... & 0.36 & 0.057 & 5.3 & ... & ... & ... \\
HIP 26369 & 2 & IC & ... & B22 & ... & 0.13 & 0.28 & 6.9 & ... & ... & ... \\
HIP 2729 & 2 & IC & ... & B22 & ... & 0.76 & 0.021 & 0.93 & ... & ... & ... \\
HIP 28036 & 2 & IC & ... & B22 & ... & 0.12 & 2.2 & 110 & ... & ... & ... \\
HIP 28764  & 3 & R & ... & G & HD 41742A & ... & 200 & 5300 & ... & ... & SB2 \\
HIP 28764  & 3 & R & ... & G & HD 41742B & ... & 200 & 5400 & ... & ... & SB2 \\
HIP 28921 & 3 & R & ... & G & Gaia DR3 2909948099477583488 & ... & 19 & 680 & ... & ... & ... \\
HIP 32938  & 3 & R & ... & G & Gaia DR3 5578901662668399872 & ... & 20 & 1100 & ... & ... & ... \\
HIP 34782 & 3 & R & ... & G & Gaia DR3 5605930265526073472 & ... & 25 & 1200 & ... & ... & PM \\
HIP 36985 & 2 & IC & ... & B22 & ... & 0.19 & 0.11 & 1.6 & ... & ... & ... \\
HIP 37349 & 3 & R & ... & G & HD 61606B & 0.65 & 58 & 820 & ... & ... & ... \\
HIP 37918 & 2 & PM, IC & ... & B22 & ... & 0.28 & 0.065 & 2.2 & ... & ... & ... \\
HIP 39826 & 3 & R & ... & G & Gaia DR3 676689831406224384 & ... & 10 & 180 & ... & ... & ... \\
HIP 43299 & 1 & RV, PM & SB1 & W07 & ... & ... & ... & ... & 0.10 & 4.0 & ... \\
HIP 44722 & 3 & R & ... & G & BD-08 2582B & ... & 8.3 & 120 & ... & ... & ... \\
HIP 49366 & 3 & R & ... & G & HD 87424B & ... & 9.7 & 230 & ... & ... & ... \\
HIP 49767 & 2 & IC & ... & B22 & ... & 0.11 & 0.29 & 13 & ... & ... & ... \\
HIP 53771  & 3 & R & ... & G & Gaia DR3 5359955053246566144 & ... & 10 & 580 & ... & ... & ... \\
HIP 54477 & 2 & PM, IC & ... & B22 & ... & 0.25 & 0.014 & 0.76 & ... & ... & ... \\
HIP 55334 & 2 & IC & ... & B22 & ... & 0.43 & 0.13 & 12 & ... & ... & ... \\
HIP 55899 & 3 & R & ... & G & CD-39 7118B & ... & 22 & 2900 & ... & ... & ... \\
HIP 55899 & 1 & PM & ... & TW & ... & ... & ... & ... & 0.080 & 11 & ... \\
HIP 56128 & 2 & IC & ... & B22 & ... & 0.34 & 0.21 & 7.1 & ... & ... & ... \\
HIP 56963 & 2 & PM, IC & ... & B22 & ... & 0.49 & 0.20 & 27 & ... & ... & ... \\
HIP 57013 & 1 & RV & SB1 & TW & ... & 0.43 & ... & ... & 0.0010 & 0.064 & ... \\
HIP 57013 & 3 & R & ... & G & Gaia DR3 5379417091947347840 & ... & 8.4 & 540 & ... & ... & ... \\
HIP 58146 & 1 & PM, NSS, RV & SB1 & GS1 & ... & ... & ... & ... & 0.080 & 8.5 & ... \\
HIP 58465 & 1 & PM & ... & TW & ... & ... & ... & ... & 0.070 & 7.0 & ... \\
HIP 58859 & 2 & IC & ... & B22 & ... & 0.20 & 0.28 & 34 & ... & ... & ... \\
HIP 59505 & 1 & PM & ... & TW & ... & ... & ... & ... & 0.050 & 6.5 & ... \\
HIP 59603 & 2 & IC & ... & B22 & ... & 0.30 & 0.088 & 10 & ... & ... & ... \\
HIP 60459 & 1 & PM & ... & TW & ... & ... & ... & ... & 0.080 & 8.2 & ... \\
HIP 60577 & 2 & IC & ... & B22 & ... & 0.69 & 2.5 & 300 & ... & ... & ... \\
HIP 60577 & 2 & IC & ... & B22 & ... & 0.69 & 2.6 & 310 & ... & ... & ... \\
HIP 60965 & 3 & R & ... & G & HD 183870B & ... & 23 & 600 & ... & ... & ... \\
HIP 61049 & 1 & PM, NSS, RV & SB1 & GS1 & ... & ... & ... & ... & 0.090 & 9.3 & ... \\
HIP 61087 & 2 & IC & ... & B22, a7 & ... & 0.18 & 0.054 & 5.8 & ... & ... & ... \\
HIP 61796 & 2 & IC & ... & B22 & ... & 0.32 & 0.26 & 29 & ... & ... & ... \\
HIP 62171 & 2 & IC & ... & B22 & ... & 0.37 & 0.13 & 15 & ... & ... & ... \\
HIP 62428 & 1 & PM, IC & ... & B22 & ... & 1.5 & 0.036 & 4.0 & 0.070 & 7.8 & ... \\
HIP 63041 & 2 & IC & ... & B22 & ... & 0.17 & 0.054 & 5.4 & ... & ... & ... \\
HIP 64322 & 1 & PM, RV & SB1 & B22, G23 & ... & ... & ... & ... & 0.10 & 10 & ... \\
HIP 64322 & 2 & IC, R & ... & B22 & Gaia DR3 5863295052504240128 & ... & 2.3 & 230 & ... & ... & ... \\
HIP 65219 & 2 & PM, IC & ... & B22 & ... & 1.4 & 0.067 & 8.0 & ... & ... & ... \\
HIP 66651 & 1 & PM & ... & TW & ... & ... & ... & ... & 0.10 & 14 & ... \\
HIP 66722 & 1 & PM & ... & TW & ... & ... & ... & ... & 0.10 & 12 & ... \\
HIP 66908 & 2 & IC & ... & B22 & ... & 0.36 & 0.15 & 14 & ... & ... & ... \\
HIP 67036 & 2 & IC & ... & B22 & ... & 0.28 & 0.29 & 38 & ... & ... & ... \\
HIP 67036 & 2 & IC & ... & B22 & ... & 0.40 & 0.31 & 42 & ... & ... & ... \\
HIP 70350 & 2 & IC & ... & B22 & ... & 1.3 & 0.25 & 30 & ... & ... & ... \\
HIP 70697 & 2 & IC & ... & B22 & ... & 0.51 & 0.58 & 78 & ... & ... & ... \\
HIP 70833 & 2 & IC & ... & B22 & ... & 0.77 & 2.9 & 730 & ... & ... & ... \\
HIP 70833 & 1 & PM, NSS & ... & a7 & ... & ... & ... & ... & 0.10 & 25 & ... \\
HIP 71321 & 2 & PM, IC & ... & B22 & ... & 0.62 & 0.15 & 19 & ... & ... & ... \\
HIP 71724 & 1 & PM & ... & TW & ... & ... & ... & ... & 0.10 & 15 & ... \\
HIP 72192 & 1 & PM & ... & TW & ... & ... & ... & ... & 0.10 & 12 & ... \\
HIP 72622 & 1 & RV, IC & SB2 & F14, W23 & ... & 1.8 & 0.020 & 0.46 & ... & ... & ... \\
HIP 72622 & 3 & R & ... & G & HIP 72603 & ... & 230 & 5400 & ... & ... & SB \\
HIP 73913 & 2 & PM, IC & ... & B22 & ... & 0.59 & 0.089 & 14 & ... & ... & ... \\
HIP 75367 & 2 & IC & ... & B22 & ... & 0.24 & 0.86 & 120 & ... & ... & ... \\
HIP 77388 & 2 & IC & ... & B22 & ... & 0.60 & 1.2 & 160 & ... & ... & ... \\
HIP 77813 & 2 & PM, IC & ... & B22 & ... & 0.82 & 0.047 & 8.3 & ... & ... & ... \\
HIP 78581 & 2 & IC & ... & B22 & ... & 0.20 & 2.4 & 250 & ... & ... & ... \\
HIP 78581 & 2 & PM, IC & ... & B22 & ... & 0.19 & 0.052 & 5.4 & ... & ... & ... \\
HIP 78754 & 1 & PM & ... & TW & ... & ... & ... & ... & 0.060 & 11 & ... \\
HIP 79031 & 1 & PM & ... & TW & ... & ... & ... & ... & 0.040 & 6.4 & ... \\
HIP 79124 & 2 & IC & ... & B22 & ... & 0.15 & 0.95 & 130 & ... & ... & ... \\
HIP 79124 & 2 & IC & ... & B22 & ... & 0.42 & 0.17 & 23 & ... & ... & ... \\
HIP 79156 & 2 & IC & ... & B22 & ... & 0.41 & 0.87 & 130 & ... & ... & ... \\
HIP 79258 & 3 & R & ... & G & Gaia DR3 6035727063034449408 & ... & 22 & 3200 & ... & ... & ... \\
HIP 82688 & 2 & IC & ... & B22 & ... & 0.21 & 3.8 & 170 & ... & ... & ... \\
HIP 85038 & 3 & R & ... & G & HD 156751B  & ... & 9.4 & 640 & ... & ... & PM \\
HIP 87386 & 2 & PM, IC & ... & B22 & ... & 0.81 & 0.18 & 12 & ... & ... & ... \\
HIP 88694 & 3 & R & ... & G & Gaia DR3 4038724053950164096 & ... & 12 & 210 & ... & ... & ... \\
HIP 93580 & 2 & PM, IC & ... & B22 & ... & 0.51 & 0.26 & 15 & ... & ... & ... \\
HIP 95149 & 2 & IC & ... & B22 & ... & 0.26 & 0.21 & 4.2 & ... & ... & ... \\
HIP 96085 & 3 & R & ... & G & Gaia DR3 4187836934609063552 & ... & 210 & 3700 & ... & ... & ... \\
HIP 97255 & 2 & IC & ... & B22 & ... & 0.29 & 0.32 & 9.4 & ... & ... & ... \\
LP 876-10 & 3 & R & ... & G & Fomalhaut B & 0.74 & 27000 & 210000 & ... & ... & ... \\
LP 876-10 & 3 & R & ... & G & Fomalhaut A & 1.9 & 20000 & 150000 & ... & ... & ... \\
TWA 11B & 3 & R & ... & G & HD 109573A & ... & 7.8 & 550 & ... & ... & ... \\
TWA 24 & 2 & IC & ... & B22 & ... & 0.28 & 3.5 & 370 & ... & ... & ... \\
TWA 5 & 1 & RV & SB1 & K13 & ... & 0.26 & 0.064 & 3.2 & 0.10 & 5.0 & ... \\
TYC 0523-0573-1 & 1 & NSS & ... & aS1 & ... & ... & ... & ... & 0.010 & 0.37 & ... \\
TYC 4895-1137-1 & 2 & PM, IC & ... & B22 & ... & 0.67 & 0.28 & 16 & ... & ... & ... \\
TYC 5164-0567-1 & 2 & IC & ... & B22 & ... & 0.43 & 2.7 & 180 & ... & ... & ... \\
TYC 6004-2114-1 & 2 & IC & ... & B22 & ... & 0.74 & 0.072 & 8.1 & ... & ... & ... \\
TYC 6299-2608-1 & 2 & IC & ... & B22 & ... & 0.65 & 0.16 & 12 & ... & ... & ... \\
TYC 6820-0223-1 & 2 & IC & ... & B22 & ... & 0.92 & 0.12 & 10 & ... & ... & ... \\
TYC 6872-1011-1 & 2 & IC & ... & B22 & ... & 0.19 & 0.041 & 3.0 & ... & ... & ... \\
TYC 7059-1111-1 & 2 & IC & ... & B22 & ... & 0.69 & 1.0 & 64 & ... & ... & ... \\
TYC 7079-0068-1 & 2 & IC & ... & B22 & ... & 0.21 & 0.088 & 7.7 & ... & ... & ... \\
TYC 7079-0068-1 & 2 & IC & ... & B22 & ... & 0.49 & 0.56 & 49 & ... & ... & ... \\
TYC 7080-0147-1 & 2 & PM, IC & ... & B22 & ... & 0.59 & 0.062 & 5.1 & ... & ... & ... \\
TYC 7133-2511-1 & 1 & RV, PM & ... & B22 & ... & 0.77 & ... & ... & 0.0017 & 0.84 & ... \\
TYC 7133-2511-1 & 2 & IC & ... & B22 & ... & 0.77 & 0.11 & 54 & ... & ... & ... \\
TYC 7191-0707-1 & 2 & IC & ... & B22 & ... & 0.91 & 0.99 & 120 & ... & ... & ... \\
TYC 7364-0911-1 & 2 & IC & ... & B22 & ... & 0.15 & 0.093 & 8.7 & ... & ... & ... \\
TYC 7379-279-1 & 2 & IC & ... & B22 & ... & 0.24 & 2.0 & 110 & ... & ... & ... \\
TYC 7379-279-1 & 2 & IC & ... & B22 & ... & 0.55 & 2.4 & 130 & ... & ... & ... \\
TYC 7627-2190-1 & 2 & IC & ... & B22 & ... & 0.70 & 0.054 & 6.3 & ... & ... & ... \\
TYC 7657-1711-1 & 2 & PM, IC & ... & B22 & ... & 0.72 & 0.076 & 7.2 & ... & ... & ... \\
TYC 7722-0207-1 & 1 & RV & SB2 & T20 & ... & 0.79 & ... & ... & 0.10 & 6.4 & ... \\
TYC 8137-2609-1 & 1 & OA & ... & TW & ... & ... & ... & ... & 0.10 & 41 & ... \\
TYC 8182-1315-1 & 2 & IC & ... & B22 & ... & 0.77 & 0.17 & 15 & ... & ... & ... \\
TYC 8332-2024-1 & 2 & IC & ... & B22 & ... & 0.91 & 0.086 & 9.6 & ... & ... & ... \\
TYC 8400-0567-1 & 2 & PM, IC & ... & B22 & ... & 0.59 & 0.062 & 3.1 & ... & ... & ... \\
TYC 8491-0656-1 & 2 & IC & ... & B22 & ... & 0.72 & 0.045 & 2.0 & ... & ... & ... \\
TYC 8491-1376-1 & 1 & NSS & ... & orb & ... & ... & ... & ... & 0.010 & 0.69 & ... \\
TYC 8497-0995-1 & 2 & IC & ... & B22 & ... & 0.48 & 0.063 & 3.1 & ... & ... & ... \\
TYC 8577-1672-1 & 2 & PM, IC & ... & B22 & ... & 0.81 & 0.082 & 8.0 & ... & ... & ... \\
TYC 8582-1705-1 & 2 & IC & ... & B22 & ... & 0.18 & 0.083 & 4.6 & ... & ... & ... \\
TYC 8634-1393-1  & 3 & R & ... & G & Gaia DR3 5343936233991962880 & ... & 38 & 1600 & ... & ... & PM \\
TYC 8911-2430-1 & 2 & IC & ... & B22 & ... & 0.42 & 0.17 & 17 & ... & ... & ... \\
TYC 8944-1516-1 & 2 & IC & ... & B22 & ... & 0.78 & 0.024 & 2.8 & ... & ... & ... \\
TYC 9493-0838-1 & 2 & IC & ... & B22 & ... & 0.43 & 0.58 & 40 & ... & ... & ... \\
UCAC2 06727592 & 2 & IC & ... & B22 & ... & 0.65 & 1.3 & 150 & ... & ... & ... \\
V4046 Sgr & 1 & RV & SB2 & A15 & ... & 0.88 & ... & ... & 0.10 & 7.1 & ... \\
V4046 Sgr & 3 & R & ... & A15 & Gaia DR3 4045698732855626624 & 0.71 & 170 & 12000 & ... & ... & ... \\

\hline

\end{longtable} 	
\tablefoot{M: type of companion (1: unresolved in SPHERE; 2: resolved in SPHERE; 3: outside field of view). Method: detection method (R: resolved by Gaia; IC: resolved in DI or interferometry; RV: radial velocities; NSS: GaiaNSS; PM: GaiaPMEX; OA: other astrometric). References: TW = this work; G = Gaia DR3; GS1 = GaiaNSS (SB1); a7 = GaiaNSS (acceleration7); a9 = GaiaNSS (acceleration9); aS1  = GaiaNSS (AstroSpectroSB1); orb = GaiaNSS (Orbital); 
ots = GaiaNSS (OrbitalTargetedSearchValidated); A15 = \citet{alonsofloriano15}; B22 = \citet{bonavita22}; B22b = \citet{bonavita22b}; F14 = \citet{fuhrmann14}; F20 = \citet{flagg20}; G23 = \citet{grandjean23}; K13 = \citet{koehler13K}; L09 = \citet{lagrange09}; L16 = \citet{lagrange16}; M14 = \citet{marion14};
T20 = \citet{traven20}; T22 = \citet{tokovinin22}; W07 = \citet{white07};
W15 = \citet{wagner15}; W23 = \citet{waisberg23}. N1, N2: notes about the primary and the companion, respectively (SB = spectroscopic binary; PM = PMEX binary, only reported here for companions).}
\end{landscape}

\end{appendix}

\end{document}